\documentclass[aps,prd,twocolumn,superscriptaddress,notitlepage,nofootinbib,preprintnumbers]{revtex4-2}

\usepackage{epsfig}
\usepackage{multirow}
\usepackage{slashed}
\usepackage{amsmath}
\usepackage{physics}
\usepackage{qcircuit}
\usepackage{braket}
\usepackage{graphicx}
\usepackage{subfigure}
\usepackage{ulem}
\usepackage{color}
\usepackage{epstopdf}
\usepackage[inline]{enumitem}

\def\bea#1\eea{\begin{align}#1\end{align}}
\newcommand{\nn}{\nonumber\\}
\newcommand{\bef}{\begin{figure}[!htp]}
\newcommand{\eef}{\end{figure}}

\begin{document}
\preprint{RIKEN-iTHEMS-Report-25}
\title{Quantum simulations of quantum electrodynamics in Coulomb gauge}

\date{\today}

\author{Tianyin Li}
\email{tianyinli@m.scnu.edu.cn}
\email{tianyin.li@riken.jp}
\affiliation{Key Laboratory of Atomic and Subatomic Structure and Quantum Control (MOE), Guangdong Basic Research Center of Excellence for Structure and Fundamental Interactions of Matter, Institute of Quantum Matter, South China Normal University, Guangzhou 510006, China}
\affiliation{Guangdong-Hong Kong Joint Laboratory of Quantum Matter, Guangdong Provincial Key Laboratory of Nuclear Science, Southern Nuclear Science Computing Center, South China Normal University, Guangzhou 510006, China}
\affiliation{RIKEN Center for Interdisciplinary Theoretical and Mathematical Sciences (iTHEMS), RIKEN, Wako 351-0198, Japan}

\begin{abstract}
In recent years, the quantum computing method has been used to address the sign problem in traditional Monte Carlo lattice gauge theory (LGT) simulations. We propose that the Coulomb gauge (CG) should be used in quantum simulations of LGT. Since the redundant degrees of freedom of gauge fields can be eliminated in CG, the Hamiltonian in CG does not need to be gauge invariance, allowing the gauge field to be discretized naively. Then the discretized gauge fields and fermion fields should be placed on momentum and position lattices, respectively. Under this scheme, the CG condition and Gauss's law can be conveniently preserved by solving for the polarization vectors from algebraic equations. Furthermore, we discuss the mapping of gauge fields to qubits and evaluate the associated qubit and gate cost of this framework. We point out that this formalism is efficient for simulating hadron scattering processes on future fault-tolerant quantum computers. Finally, we calculate the vacuum expectation value of the U(1) plaquette operator and the Wilson loop on a classical device to test the performance of our discretization scheme.
\end{abstract}

\maketitle

\section{Introduction}
Gauge field theory is an essential cornerstone of the standard model (SM) as it describes the fundamental laws governing the interactions between elementary particles \cite{Yang:1954ek,Peskin:1995ev,Weinberg:1995mt}. There are non-perturbative problems within the SM, such as the long-distance physics in Quantum Chromodynamics (QCD), including the QCD phase diagram \cite{Shuryak:1980tp}, hadron structure \cite{Collins:1989gx}, etc. Non-perturbative approaches, such as lattice QCD, have been proposed to address these problems \cite{Wilson:1974sk,Ding:2015ona,Ji:2020ect}. However, due to the sign problem of the Monte Carlo algorithms \cite{Troyer:2004ge}, lattice QCD faces challenges when simulating time evolutions and finite-density systems. These challenges prompt the exploration of alternative non-perturbative methods for simulating gauge field theories. The quantum computing method is one of the choices \cite{Feynman:1981tf}.

The quantum simulations of gauge fields involve a complete set of procedures \cite{Byrnes:2005qx,Lamm:2019bik,Bauer:2022hpo,Bauer:2023qgm}. The first step is the discretization of gauge and fermion fields, with the Kogut-Susskind (K-S) formalism being the most widely employed discretization scheme \cite{Kogut:1974ag}. This formalism is based on choosing a temporal gauge, $A^0=0$, where not all residual degrees of freedom are eliminated. Consequently, the K-S Hamiltonian must remain invariant under residual gauge transformations. Digitizing the continuous gauge group is the second step of simulating lattice gauge theory (LGT) on quantum computers. A critical aspect of this step is the preservation of Gauss's law, which ensures gauge invariance. In (1+1)-dimensional cases, this can typically be achieved, as the gauge field can be eliminated by solving Gauss's law \cite{Klco:2018kyo,Shaw:2020udc,Gustafson:2020yfe,Florio:2023dke,Davoudi:2024wyv,Farrell:2024fit,deJong:2021wsd,Xie:2022jgj,Ikeda:2023zil,Lee:2023urk,Dempsey:2023gib,Atas:2021ext,Farrell:2022wyt,Farrell:2022vyh,Zhang:2024fgv}. However, maintaining Gauss's law in higher dimensions is more complex. Several methods have been proposed to address this challenge, such as the truncation of the electric field (group representation) basis \cite{Byrnes:2005qx,Zohar:2012xf,Zohar:2012ay,Zohar:2013zla,Klco:2019evd,Davoudi:2020yln,Paulson:2020zjd,ARahman:2021ktn,Ciavarella:2021nmj,Ciavarella:2024fzw}, discrete subgroup method \cite{Zohar:2014qma,Ji:2020kjk,Haase:2020kaj,Alexandru:2021jpm,Carena:2021ltu,Armon:2021uqr,Gustafson:2022xdt,Gustafson:2023kvd,Irmejs:2022gwv,Hartung:2022hoz,Charles:2023zbl,Carena:2024dzu,Lamm:2024jnl}, futher gauge fixing \cite{DAndrea:2023qnr,Grabowska:2024emw,Grabowska:2024vop,Kreshchuk:2020aiq,Kreshchuk:2020dla,Kreshchuk:2020kcz,Farrell:2022wyt,Farrell:2022vyh}, Loop-string-hadron formalism \cite{Raychowdhury:2018osk,Raychowdhury:2019iki,Kadam:2022ipf,Kadam:2023gpn,Kadam:2024zkj} and the quantum link model \cite{Luo:2019vmi,Brower:2020huh,Mathis:2020fuo}.

%Though the K-S formalism has achieved great success, this discretization scheme adopts the temporal gauge, wherein Gauss’s law becomes a constraint on the state space rather than a direct constraint on the field operators. This makes a part of the Hilbert space redundant and requires the Hamiltonian to be invariant under the residual gauge transformation. As we know, the dimension of the Hilbert space on a gauge link is infinite due to the infinite number of group elements of the continuous Lie group. Since the Hilbert space of a quantum computer is finite, it is necessary to truncate the group element basis or representation basis on the gauge link. These truncations will break Gauss’s law, which guarantees gauge invariance. The breaking of Gauss’s law implies that the Gauss’s law operator does not commute with the Hamiltonian, so the Hamiltonian and Gauss’s law operators do not have common eigenstates. However, the hadron states of QCD are color-neutral Hamiltonian eigenstates, which are the common eigenstates of Gauss’s law operator and the Hamiltonian. To make the hadron state well-defined, Gauss’s law needs to be preserved on the lattice.

Though the K-S formalism has achieved great success, it has challenges when using this formalism to simulate the hadron scattering process in more than $(1+1)$-dimensional space-time on quantum computers. According to the factorization theorem \cite{Collins:1989gx}, the cross-section of high energy hadron scattering $\sigma$ can be factorized into $\sigma=f\Tilde{\otimes} \hat{\sigma} \Tilde{\otimes} D$, where $\Tilde{\otimes}$ denotes a convolution operator. In this expression, $f$ represents parton distribution functions, $\hat{\sigma}$ is the hard cross-section for quarks and gluons, and $D$ denotes non-perturbative functions that characterize the hadronization of partons. The calculation of those quantities has been explored in \cite{Ji:2013dva,Ji:2019sxk,Lamm:2019uyc,Li:2022lyt,Li:2023kex,Li:2024nod}. Ideally, we aim to calculate $f$, $\hat{\sigma}$, and $D$ using the same formalism to avoid the complications associated with matching results obtained from different methods. This desired formalism should meet the following requirements: 1) It must efficiently reach the continuum limit. 
%2). It should make the hadron state well-defined after digitizing the gauge group. 
2) The Wilson line must be unitary, as the operator definitions of $f$ and $D$ involve Wilson lines. 3) The final state of $\hat{\sigma}$ may include polarized gluons or photons, necessitating straightforward handling of these states within the formalism. 4). When simulating inclusive processes, it is crucial to sum over all physical asymptotic ``out" states, which are the eigenstates of the Hamiltonian $H$. Thus, it is better to preserve Gauss's law $G\ket{\rm Phys} = 0$ as well as $[G,H] = 0$ strictly after digitizing the gauge fields. This guarantees the gauge invariance of all ``out" states. However, those requirements may be sacrificed if we don't digitize the gauge fields carefully. For example, truncating the electric basis to the maximum eigenvalue $\Lambda$ will break the commutation relation $[G,H] = 0$ at order $O(1/\Lambda)$. This implies that Gauss’s law operator does not commute with the Hamiltonian, so the Hamiltonian and Gauss’s law operators do not have common eigenstates. The quantum link model preserves Gauss's law while truncating the electric basis, but this method sacrifices the unitarity of the gauge link. The discrete subgroup method can preserve Gauss’s law after digitizing the gauge group, and it can also reach the continuum limit of some model \cite{Bhattacharya:2020gpm}. However, the largest subgroup of SU(3) is S(1080), so this method might sacrifice requirement 1) in the SU(3) gauge theory. We can do further gauge fixing by solving Gauss's law on the lattice. However, this method has a challenge when considering the theory with dynamical quarks, which is necessary to define the hadron.

In this work, we propose a formalism for simulating LGT on a quantum computer that aims to satisfy all four requirements, at least in the case of Quantum Electrodynamics (QED). At a cost, we need to introduce non-local interaction terms in the Hamiltonian, so more quantum gates are needed when simulating time evolution. Different from K-S formalism, our formalism is based on fixing the Coulomb Gauge (CG). The choice of CG is motivated by its preservation of rotational symmetries, unlike the axial and light-cone gauges. Compared with the maximum tree gauge, CG can handle dynamical fermions naturally. Once the CG is fixed, there are no nontrivial gauge transformations that can simultaneously preserve both the Hamiltonian and the constraints. From this perspective, we can utilize the discrete version of $A_\mu(x)$ to construct the Hamiltonian together with constraints $\chi(A)=0$ (such as Gauss's law and the CG condition) on the lattice (see supplementary materials). The lattice setup will be discussed in Section \ref{sec:dis_cgqed}. After that, the mapping of gauge fields $a^{r}_\mathbf{p}$ and the fermion fields $\psi_\alpha(\mathbf{x})$ to qubits will be given in section \ref{sec:map_to_qb}. Sections \ref{sec:compliexity} and \ref{sec:results} address the complexity of our framework and present our results, respectively.

%In this formalism, requirement 1) is satisfied because the continuum limit corresponds to the weak coupling limit. In this limit, the field operators only create or annihilate a small number of photons, allowing for the continuum limit to be approached with a relatively modest truncation of the photon Fock states. Requirement 2) is inherently satisfied, as Gauss's law with dynamical fermions can also be resolved within the CG. Additionally, requirement 3) is fulfilled due to the presence of a Hermitian $A_\mu$, which can generate a unitary Wilson line. Requirement 5) is addressed since all residual degrees of freedom are eliminated in the CG, which also implies that fewer qubits are necessary to encode the gauge fields.

\begin{figure*}
    \centering
    \includegraphics[width=0.98\linewidth]{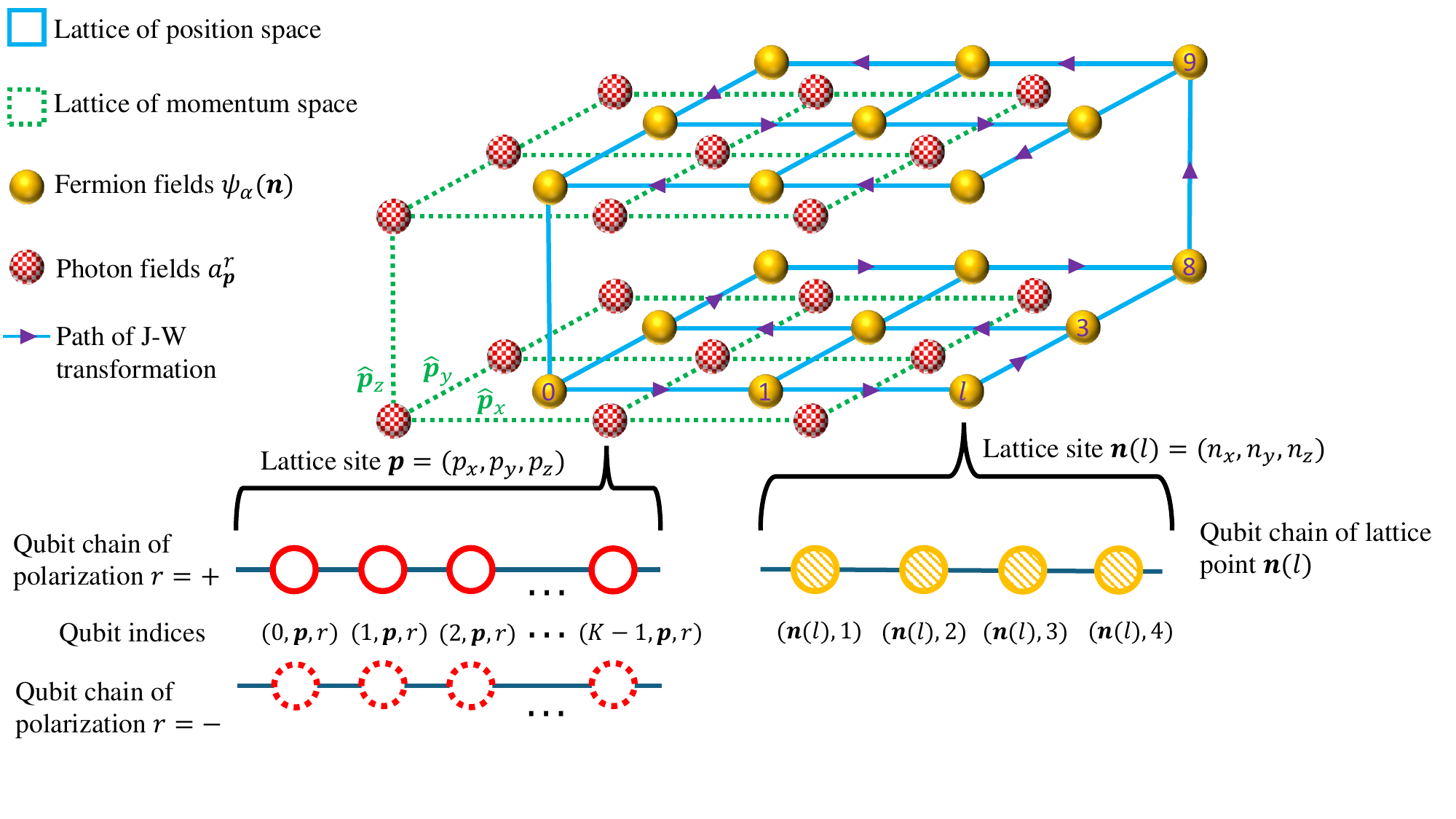}
    \caption{The (3+1)-dimensional lattice of position space (blue) and momentum space (green) is depicted. Fermion fields $\psi_\alpha(\mathbf{n})$ and photon fields $a^r_\mathbf{p}$ are placed on the position space lattice and the momentum space lattice, respectively. Under the choice of Fock state truncation $\Lambda = 2^K - 1$, $2K$ qubits need to be used to represent the photon fields at each point $\mathbf{p}$ on the momentum lattice. These $2K$ qubits are labeled by three indices $(J, \mathbf{p}, r)$, where $J = 0, 1, 2, ..., K - 1$. Four qubits, labeled by $(\mathbf{n}, \alpha)$, are used to represent the fermion fields $\psi_\alpha(\mathbf{n})$ at lattice point $\mathbf{n}$. The purple arrows show the path of the Jordan-Wigner transformation, while the purple numbers $0, 1, ..., l, ...$ are the parameters of the path.}
    \label{fig:lat_set}
\end{figure*}

\section{Discretizations of Coulomb gauge QED}\label{sec:dis_cgqed}
    Both fermion fields and gauge fields need to be discretized. Here, we give important results of CG QED on the lattice, while the details can be found in the supplementary materials. As shown in Fig.~\ref{fig:lat_set}, we put the fermion fields $\psi_\alpha(\mathbf{n})$ and gauge fields $a^r_{\mathbf{p}}$ in the lattice of position and momentum space, respectively, where $a^r_{\mathbf{p}}$ is the annihilation operator of photons with given momentum $\mathbf{p}$ and porlization $r=\pm$. On the lattice, the gauge potential $\hat{A}_0(\mathbf{n})$, $\hat{A}_i(\mathbf{n})$ and its conjugate momentum $\hat{\Pi}_{\perp,i}$ can be represented by:
    \bea\label{eq:disc}
        &\hat{A}^0(\mathbf{n}) = \sum_{\mathbf{m}} \sum_{\mathbf{p}\not=0} \frac{J^0(\mathbf{m})}{M^3\hat{E}_{\mathbf{p}}^2}e^{-i\mathbf{p}\cdot(\mathbf{n}-\mathbf{m})}\,,\nonumber\\
        &\hat{A}_i (\mathbf{n}) = \sum_{\mathbf{p}\not= \mathbf{0}} \frac{1}{\sqrt{2\hat{E}_{\mathbf{p}} M^3}} \sum_{r} \left[\hat{\epsilon}^r_i (\mathbf{p}) a^r_\mathbf{p} e^{i\mathbf{p}\cdot \mathbf{n}}+{\rm H.c.}\right]\,,\nn
        &\hat{\Pi}_{\perp i} (\mathbf{n}) = \sum_{\mathbf{p}\not = \mathbf{0}} \sqrt{\frac{\hat{E}_{\mathbf{p}}}{2M^3}} \sum_{r} \left[-i\hat{\epsilon}^r_i (\mathbf{p}) a^r_\mathbf{p} e^{i\mathbf{p}\cdot \mathbf{n}}+{\rm H.c.}\right]\,,
    \eea
    where $J^\mu(\mathbf{n})=g\bar{\psi}(\mathbf{n})\gamma^\mu \psi(\mathbf{n})$ is the electric current, $\hat{E}_{\mathbf{p}} \equiv \sqrt{\frac{4}{a^2}\sum_{i} \sin^2 \left(\frac{p^i a}{2}\right)}$ is the dispersion relation of photon on the lattice with lattice spacing $a$, and $M$ is the number of lattice site in each special dimension. On the lattice of momentum space, the momentum $\mathbf{p}$ can be evaluated as $\mathbf{p} = (2\pi k_x/Ma,2\pi k_y/Ma,2\pi k_z/Ma)$, with $k_x,k_y,k_z=0,\pm1,...,\pm \lfloor M/2 \rfloor$. The unitary gauge link, given by $U_i(\mathbf{n})=\exp(-igA_i(\mathbf{n}))$, fulfills the requirement 2) in the introduction. $\hat{\epsilon}^r_i(\mathbf{p})$ in the Eq.~(\ref{eq:disc}), are polarization vectors of photon, which need to be determined by solving:
    \bea\label{eq:lateps}
        &\sum_i (e^{i p^i a}-1) \hat{\epsilon}^r_i(\mathbf{p}) = 0\,, \nn
        &\sum_i \hat{\epsilon}^r_i(\mathbf{p}) \hat{\epsilon}^{s}_i(\mathbf{p}) = \delta_{rs}\,,\nn
        &\sum_{r} \hat{\epsilon}^r_i(\mathbf{p}) \hat{\epsilon}^r_j(\mathbf{p}) = \delta_{ij} -\frac{1}{\hat{E}_{\mathbf{p}}^2} (e^{-i p^i a}-1) (e^{i p^j a}-1)\,.
    \eea
    Once the solution of $\hat{\epsilon}^r_i(\mathbf{p})$ is obtained, Gauss's law and CG condition will be satisfied automatically. It's worth noting that both $\hat{A}_0$ and $\hat{\epsilon}^r_i(\mathbf{p})$ do not depend on the truncation of photon Fock states. This means that all constraints, including Gauss's law and the CG condition, will be preserved accurately after the digitization of gauge fields, which is important for simulating gauge field theories on quantum computers, and fulfill requirement 4) in our introduction.

    After discretizations of fields, the Hamiltonian of CG QED is given by $\hat{H} = \hat{H}_E+\hat{H}_B+\hat{H}_I+\hat{H}_V+\hat{H}_M+\hat{H}_W$ (we set $a=1$ here):
    \bea\label{eq:dis_ham}
        &\hat{H}_E + \hat{H}_B = \sum_{\mathbf{p}\not = \mathbf{0}} \sum_{r} \hat{E}_{\mathbf{p}} a^{r\dagger}_{\mathbf{p}} a^{r}_{\mathbf{p}}\,,\nn
        &\hat{H}_I = \sum_{\mathbf{n},i} \sum_{\mathbf{p}\not=0} \sum_{r} \frac{J^i(\mathbf{n})}{M^{\frac{3}{2}}\sqrt{2\hat{E}_{\mathbf{p}}}}\left[\hat{\epsilon}^r_i (\mathbf{p}) a^r_\mathbf{p} e^{i\mathbf{p}\cdot \mathbf{n}}+{\rm H.c.}\right]\,,\nn
        &\hat{H}_V = \frac{1}{2}\sum_{\mathbf{n},\mathbf{m}} \sum_{\mathbf{p}\not=0} \frac{J^0(\mathbf{n})J^0(\mathbf{m})}{\hat{E}_{\mathbf{p}}^2}e^{-i\mathbf{p}\cdot(\mathbf{n}-\mathbf{m})}\,,\nn
        &\hat{H}_M = \sum_{\mathbf{n},i} \bar{\psi}(\mathbf{x})\left[-i \gamma^i \frac{\psi(\mathbf{n}+\hat{i})-\psi(\mathbf{n}-\hat{i})}{2}+m\bar{\psi}\psi\right]\,,\nn
        &\hat{H}_W = \sum_{\mathbf{n}}-\frac{w}{2} \bar{\psi}(\mathbf{n}) \hat{\Delta} \psi(\mathbf{n})\,,
    \eea
    where $\hat{H}_W$ comes from discretizing the fermion field by using the Wilson fermion method, and the parameter $w$ needs to satisfy $0<w<1$.

\section{Mapping fields to qubits}\label{sec:map_to_qb}
    Both fermion fields $\hat{\psi}(\mathbf{n})$ and photon annihilation operators $a^r_{\mathbf{p}}$ need to be mapped to qubits. In Fig.~(\ref{fig:lat_set}), we assume $K$ qubits are used to represent Hilbert space $\mathcal{H}^{\rm ph}_{\mathbf{p},r}$ of photons with momentum $\mathbf{p}$ and helicity $r$. Then, the entire Hilbert space of photons can be written as $\mathcal{H}^{\rm ph}=\otimes_{\mathbf{p},r}\mathcal{H}_{\mathbf{p},r}$, where $\otimes$ denotes the direct product. The computational basis of $\mathcal{H}^{\rm ph}_{\mathbf{p},r}$ can be written as $\ket{\mathcal{N}}_{\mathbf{p},r} \equiv \ket{i_0i_1...i_{K-1}}_{\mathbf{p},r}$, where $\mathcal{N}$ is the decimal numeral corresponding to the binary number $i_0i_1,...,i_{K-1}$, which also can be denoted as $\mathcal{N}(i_{K-1},...,i_{0})$. As shown in Fig.~\ref{fig:lat_set}, three lower indices $(J,\mathbf{p},r)\,, J=0,1,...,K-1$ are used to label the qubits in the state $\ket{\mathcal{N}}_{\mathbf{p},r}$. $a^r_\mathbf{p}$ and $a^{r\dagger}_\mathbf{p}$ can raise and low the state $\ket{\mathcal{N}}_{\mathbf{p},r}$
    \bea\label{eq:aad}
        &a^r_{\mathbf{p}} \ket{\mathcal{N}}_{\mathbf{p},r} = \sqrt{\mathcal{N}} \ket{\mathcal{N}-1}_{\mathbf{p},r}\,,\nn
        &a^{\dagger r}_{\mathbf{p}} \ket{\mathcal{N}}_{\mathbf{p},r} = \sqrt{\mathcal{N}+1} \ket{\mathcal{N}+1}_{\mathbf{p},r}\,.
    \eea
    To satisfy the Eq.~(\ref{eq:aad}) the qubit representations of $a^{r}_\mathbf{p}$ and $a^{r\dagger}_\mathbf{p}$ need to be written as:
    \bea\label{eq:map_a}
        &a^{r}_{\mathbf{p}} = \left\{\sum_{J=0}^{K-1} \left[\sigma^+_{J,\mathbf{p},r} \left(\prod_{L=0}^{J-1}\sigma^-_{L,\mathbf{p},r}\right)\right]\right\} \sqrt{\hat{\mathcal{N}}_{\mathbf{p},r}}\,,\nn
        &a^{r\dagger}_{\mathbf{p}} = \sqrt{\hat{\mathcal{N}}_{\mathbf{p},r}} \sum_{J=0}^{K-1} \left[\sigma^-_{J,\mathbf{p},r} \left(\prod_{L=0}^{J-1}\sigma^+_{L,\mathbf{p},r}\right)\right]\,.
    \eea
    where $\sigma^+=\frac{1}{2}(\sigma^1+i\sigma^2)$ and $\mathcal{\hat{N}}_{\mathbf{p},r}$ is the particle number operator of the photon with quantum numbers $(\mathbf{p},r)$:
    \bea
        \hat{\mathcal{N}}_{\mathbf{p},r} = \sum_{J=0}^{K-1} 2^{J} \left[\frac{1}{2} \left(I-\sigma^3_{J,\mathbf{p},r}\right)\right]\,.
    \eea
    Here, the Pauli operators $I$, $\sigma^x$, $\sigma^y$, and $\sigma^z$ are denoted as $\sigma^0$, $\sigma^1$, $\sigma^2$ and $\sigma^3$. In Eq.~(\ref{eq:map_a}), the Fock states with photon occupation numbers greater than $\Lambda = 2^K-1$ need to be truncated, as $a^{r\dagger}_{\mathbf{p}}\ket{\Lambda} = 0$ due to the finite-dimensional Hilbert space of a quantum computer. The square root of $\hat{\mathcal{N}}_{\mathbf{p},r}$ is given by:
    \bea\label{eq:sqrtN}
    \sqrt{\hat{\mathcal{N}}_{\mathbf{p},r}} = \sum_{\mu_0,...,\mu_{K-1}=0,3} f_{\mu_{K-1},...,\mu_0} \sigma^{\mu_{K-1}}_{{K-1},\mathbf{p},r}\otimes ... \otimes \sigma^{\mu_0}_{0,\mathbf{p},r},
    \eea
    with
    \bea\label{eq:sqrtN_coff}
        f_{\mu_{K-1},...,\mu_0} =& \frac{1}{2^K}\sum_{i_0,...,i_{K-1}=0}^1\sqrt{\mathcal{N}(i_{K-1},...,i_0)}\nonumber\\ 
        &\times (-1)^{i_{K-1}\mu_{K-1}}\times ... \times (-1)^{i_0 \mu_0}\,.
    \eea
    We point out that Eq.~(\ref{eq:sqrtN}) and Eq.~(\ref{eq:sqrtN_coff}) can be generalized to other functions $F(\mathcal{\hat{N}}_{\mathbf{p},r})$ of particle-number operator. To achieve this, we only need to replace $\sqrt{\mathcal{N}(i_{K-1},...,i_0)}$ in the Eq.~(\ref{eq:sqrtN_coff}) by $F(\mathcal{N}(i_{K-1},...,i_0))$. The successful mapping of $a^{r\dagger}_{\mathbf{p}}$ to qubits implies that we can effectively handle photons with specific polarization and momentum on quantum computers, thereby satisfying requirement 3) outlined in the introduction.
    
    The fermion fields $\hat{\psi}_\alpha(\mathbf{n})$ can be mapped to qubits by Jordan-Winger (J-W) transformation~\cite{backens_shnirman_makhlin_2019},
    \bea
        \psi_\alpha (\mathbf{n}) =& \left[\prod_{l'<l}\left(\prod_{\beta=1}^{4}\sigma^z_{\mathbf{m}(l'),\beta}\right)\right]\nn
        &\times \left(\prod_{\beta=1}^{\alpha-1}\sigma^z_{\mathbf{n}(l),\beta}\right)\times \sigma^+_{\mathbf{n}(l),\alpha}\,,
    \eea
    where $l$ is the path parameter of J-W transformation (see Fig.~\ref{fig:lat_set}). We use the lower indices $(\mathbf{n},\alpha)$ to label the qubits representing the fermion degrees of freedom. We can see that the J-W transformation is highly non-local. However, the fermion fields always appear as $\bar{\psi}\psi$ in the position space. Therefore, put the fermion field on the lattice of position space instead of momentum space can significantly reduce the non-locality of the qubit Hamiltonian.  Now, both fermion and gauge fields are mapped to the qubits and the whole Hilbert of QED can be written as $\mathcal{H} = H^{\rm ph}\otimes H^{\rm f}$, where $H^{\rm f}$ is the Hilbert of fermions.

    \begin{figure}[htbp]
        \centering
        \includegraphics[width=0.98\linewidth]{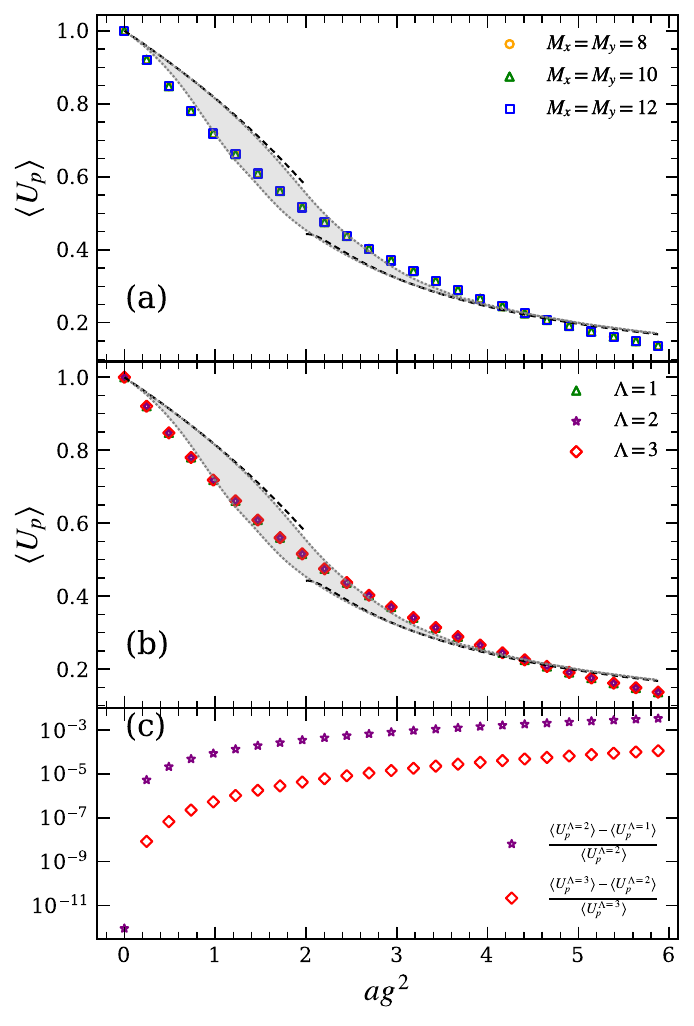}
        \caption{(a) $\braket{U_p}$ dependence on coupling constant $g$ with different lattice volumes $M_x \times M_y$ and a given Fock state cutoff $\Lambda = 1$. The points are our results, and the black dotted lines are the weak and strong coupling expansion results in \cite{Horsley:1981gj} and \cite{Balian:1974xw}. The grey region is the bootstrap allowed region that was presented in \cite{Li:2024wrd}. (b) $\braket{U_p}$ dependence on coupling constant $g$ with different Fock state cutoffs $\Lambda$ and a given lattice volume $M_x = M_y = 10$. (c) Relative errors of $\braket{U_p}$ for different cutoffs. We set $a=1$ in all of those subfigures.}
        \label{fig:Bp}
    \end{figure}

\section{Complexity}\label{sec:compliexity}
    Let's begin with the estimation of qubit scaling. Consider a $d$-dimensional spatial lattice system with $M^d$ lattice sites on $(d+1)$-dimensional space-time. Fermion fields with $N_s$ spinor components and $N_f$ flavors are discretized on the lattice system. Then, $N_f N_s M^d$ qubits need to be used to encode the fermion fields. For example, the blue lattice in Fig.~\ref{fig:lat_set} shows the case with $N_f=1$, $N_s=4$ and $d=3$. Photon fields on this lattice have $d-1$ different transverse directions. Therefore, $(d-1)\times \log_2 \Lambda \times M^d$ are needed to encode the photon fields. The case of $d-1=2$ and $\Lambda=2^K-1$ can be found in the green lattice of Fig.~\ref{fig:lat_set}. The total qubit cost should sum over both the qubit cost of encoding fermion fields and photon fields:
    \bea\label{eq:Nqcg}
        N^{\rm q,F}_{\rm CG} + N^{\rm q,G}_{\rm CG} =[N_f N_s +(d-1) \log_2 \Lambda] M^d\,,
    \eea
    where $N^{\rm q,F}_{\rm CG}$ and $N^{\rm q,G}_{\rm CG}$ are the qubit cost of encoding the fermion fields and gauge fields in our formalism. 

    In the Eq.~\eqref{eq:Nqcg}, the cutoff $\Lambda$ should be further estimated. Suppose the physics we are concerned about has an energy scale $E = \braket{\Psi|H|\Psi}$. If we want to control the Fock state truncation error of the arbitrary state $\ket{\Psi}$ under $\varepsilon_s$, the truncation $\Lambda$ should be no more than $\Lambda_{\max}$:
    \begin{align}\label{eq:bodLam}
        \Lambda &\leq \frac{(d-1) M^d}{2\varepsilon_s} \left\{\left[\frac{8 d g}{(2\pi)^{\frac{3}{2}}}  M^{\frac{d+3}{2}}+\left(\frac{8 d^2 g^2}{\pi^3} M^{d+3} \right.\right.\right. \nonumber\\ 
        &\left. \left. \left. \quad +\frac{[3 + 4 d^2 (d-1)] g^2}{8\pi^3} M^{2d+3}\right. \right. \right. \nonumber\\ 
        &\left.\left.\left. \quad + \frac{2}{\pi}\sqrt{3+m^2+12mw+36w^2} M^{d+1} \right. \right. \right. \nonumber \\ 
        &\left. \left. \left. \quad + \frac{E}{\pi} M \right)^{\frac{1}{2}}\right]^2 -1\right\} \equiv \Lambda_{\max}\,.
        \end{align}
    The proof of this inequality can be found in the supplementary materials, and the procedure of proof is similar to \cite{Jordan:2012xnu,Yao:2025uxz}. From Eq.~\eqref{eq:bodLam}, we find that $\Lambda$ depends on $M$, $g$, $E$, $m$, $w$, and $\varepsilon_s$ polynomially, which shows our formalism can reach the continuum limit of QED efficiently as well as satisfy requirement 1). The Eq.~\eqref{eq:bodLam} shows that the worst scaling behavior depends on $\varepsilon_s$ because we consider an arbitrary state $\ket{\Psi}$ in the Hilbert space. If we consider a specific case, such as the ground state of the Hamiltonian, the truncated state will converge to the untruncated state exponentially (see supplementary materials). To make Eq.~\eqref{eq:bodLam} look simpler, we can consider $m=0$ case and keep the leading term in the continuum limit $M \to \infty$:
    \begin{align}\label{eq:scLam}
        \Lambda \sim O\left(\frac{g^2M^{3d+3}+EM^{d+1}}{\varepsilon_s}\right) 
    \end{align}
    From Eq.~\eqref{eq:scLam}, we find that in the continuum limit, the $N^{\rm q,G}_{\rm CG}$ scales as
    \begin{align}
        N^{\rm q,G}_{\rm CG} &\sim (d-1) M^d \left\{\log_2\frac{1}{\varepsilon_s} \right. \nonumber\\
        &\left.\quad + \log_2\left[O\left(g^2M^{3d+3}+EM^{d+1}\right)\right]\right\}\,.
    \end{align}
    
    The complexity of simulating time evolution also needs to be estimated. For a given precision $\varepsilon$ and a Hamiltonian $H$ with a sum over $\mathcal{M}_G$ different terms, the time evolution operator $e^{-i\hat{H}t}$ can be approximated by the second-order Suzuki formula~\cite{Childs:2019hts}:
    \bea
        e^{-i\hat{H}t} \approx & \left[ \prod_{\alpha=\mathcal{M}_G}^{1} \exp\left(-i h_\alpha \frac{\delta t}{2}\right) \prod_{\alpha=1}^{\mathcal{M}_G} \exp\left(-i h_\alpha \frac{\delta t}{2}\right)\right]^{N_T} \nonumber\\
        &+\mathcal{O}\left(\varepsilon=\frac{t^3}{N_T^2}\right)\,,
    \eea
    where $\delta t = t/N_T$, and $\varepsilon$ is the error of the trotter algorithm. The $h_\alpha$ is a Pauli string multiply by a coefficient, so the evolution operator $\exp(-i h_\alpha \delta t)$ can be decomposed into one-qubit and CNOT gates (see supplementary materials). To estimate the gate cost of simulating $\exp(-i\hat{H}\delta t/2)$, we map the Hamiltonian of CG QED to qubits explicitly and give the qubit Hamiltonian in the supplementary materials. After some calculations (see supplementary materials), we obtain the one-qubit and CNOT gate cost of simulating $\exp[-i(\hat{H}_E+\hat{H}_B)\delta t/2]$ are:
    \bea
        &N^{G,EB}_{\rm OQ} = (d-1)M^d \log_2\Lambda\,, \nonumber\\
        &N^{G,EB}_{\rm CNOT} = 0\,.
    \eea
    The most complex evolution operator is $\exp(-i\hat{H}_I \delta t/2)$:
    \bea
        N^{G,I}_{\rm OQ} =& 4(d-1)M^{2d} \left[d \Lambda (\log_2 \Lambda) (\log_2\Lambda + 5) \right.\nonumber\\
        &\left.- 4\Lambda \log_2\Lambda\right]\,, \nonumber\\
        N^{G,I}_{\rm CNOT} =& 2(d-1)M^{2d} \left[d\Lambda (\log_2\Lambda) (3\log_2\Lambda+5)\right.\nonumber\\
        &\left.-4\Lambda\log_2\Lambda\right]\,.
    \eea
    Then, we need estimate the gate cost to simulate $\exp(-i\hat{H}_V \delta t/2)$:
    \bea
	&N^{G,V}_{\rm OQ} = N_s M^{2d}\,, \nonumber\\
	&N^{G,V}_{\rm CNOT} = 2N_s M^d (N_s M^d-1)\,.
    \eea
    The gate cost of simulating $\exp(-i\hat{H}_M\delta t/2)$ is
    \bea
        &N^{G,M}_{\rm OQ}(d=2) = 72 M^2\,,\nonumber\\
        &N^{G,M}_{\rm OQ}(d=3) = 104 M^3\,,\nonumber\\
        &N^{G,M}_{\rm CNOT}(d=2) = 16 N_s M^2(1+M) + 8M^2\,,\nonumber\\
        &N^{G,M}_{\rm CNOT}(d=3) = 16 N_s M^3(1+M+M^2) + 8M^3\,,
    \eea
    where $N_s=4$ both in $d=2$ and $d=3$ and we only consider the $N_f=1$ case here. The gate cost of simulating $\exp(-i\hat{H}_W \delta t/2)$ are the same as $N^{G,M}_{\rm OQ}$ and $N^{G,M}_{\rm CNOT}$. Finally, we can obtain $e^{-i\hat{H}t}$ by executing $N_T = t^\frac{3}{2}/\varepsilon^{\frac{1}{2}}$ times of $e^{-i\hat{H}\delta t}$, so the complixity of simulating $e^{-i\hat{H}t}$ is
    \bea
        N^{G}_{\rm CG,OQ} =& \frac{t^{3/2}}{\varepsilon^{1/2}}\left(N^{G,EB}_{\rm OQ}+N^{G,I}_{\rm OQ}\right. \nonumber\\
        &\left.+N^{G,V}_{\rm OQ}+N^{G,M}_{\rm OQ}+N^{G,W}_{\rm OQ}\right)\,,\nonumber\\
        N^{G}_{\rm CG,CNOT} =& \frac{t^{3/2}}{\varepsilon^{1/2}}\left(N^{G,EB}_{\rm CNOT}+N^{G,I}_{\rm CNOT}\right.\nonumber\\
        &\left.+N^{G,V}_{\rm CNOT}+N^{G,M}_{\rm CNOT}+N^{G,W}_{\rm CNOT}\right)\,.
    \eea
    The leading term of $N^{G}_{\rm CG}$ in the $M\to\infty$ and $\Lambda \to \infty$ limit is
    \begin{align}
        N^{G}_{\rm CG} \sim O\left(\frac{t^{3/2}}{\varepsilon^{1/2}} M^{2d} \Lambda (\log_2\Lambda)^2\right).
    \end{align}
    In the K-S formalism, the leading term of the gate cost for simulating $e^{-iH_{\rm KS}t}$ of U(1) gauge theory is \cite{Kan:2021xfc}:
    \begin{align}
           N^{G}_{\rm KS} \sim O\left(\frac{t^{3/2}}{\varepsilon^{1/2}}M^{3d/2}\Lambda_{\rm KS}(\log_2\Lambda_{\rm KS})^2\right)\,,
    \end{align}
    where $\Lambda_{\rm KS}$ is the Hilbert space truncation of K-S formalism in some basis, which is different from our truncation $\Lambda$.
    Compared with K-S formalism, our formalism needs an additional gate cost that is proportional to $M^{d/2}$ to simulate time evolution. This additional gate cost comes from the non-local interactions $\hat{H}_I$ in our formalism.

    \begin{figure}
        \centering
        \includegraphics[width=0.98\linewidth]{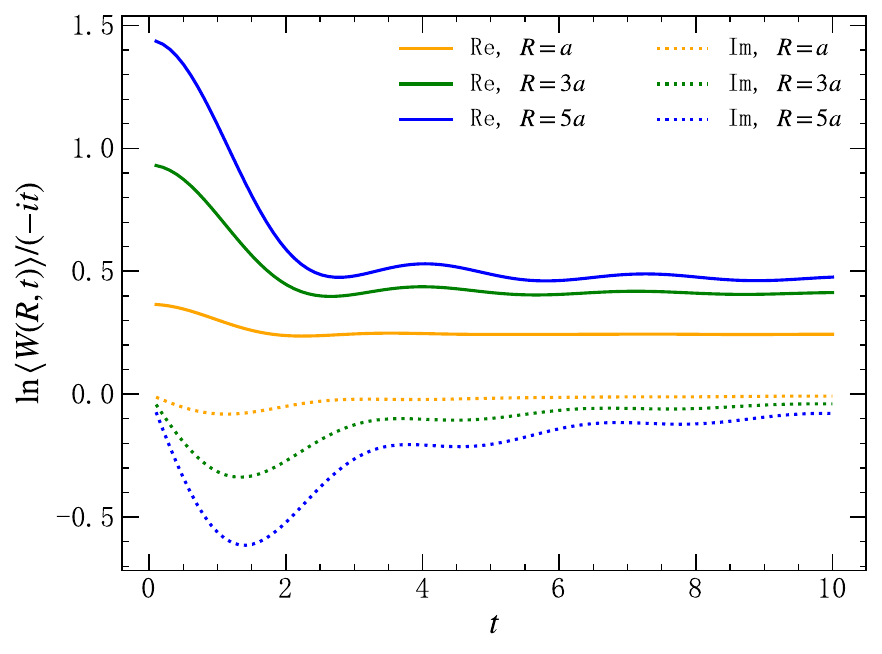}
        \caption{The real part (solid line) and imaginary part (dotted line) of $\frac{1}{-it}\ln\braket{W(R,t)}$, with $a=1$, $M_x=M_y=16$, $\Lambda=4$ and $g=0.7$.}
        \label{fig:wloop}
    \end{figure}

\section{Results}\label{sec:results}
    To test the performance of our formalism, we simulate $(2+1)$-dimensional pure U(1) gauge theory on a classical simulator. We first calculate vacuum expectation value  (VEV) of $\braket{U_p(\mathbf{n})}$, where $U_p$ is the plaquette operator of U(1) gauge theory:
    \bea
        U_p(\mathbf{n}) = U_i(\mathbf{n}) U_j(\mathbf{n}+\hat{i}) U^\dagger_i(\mathbf{n}+\hat{j}) U^\dagger_j(\mathbf{n}),
    \eea
    with $U_i(\mathbf{n}) = \exp(-ig\hat{A}_i(\mathbf{n}))$. Because of the translation invariant of the vacuum state, the VEV $\braket{U_p(\mathbf{n})}$ is independent of $\mathbf{n}$. Therefore, the $\braket{U_p(\mathbf{n})}$ can be written as $\braket{U_p}$. In Fig.~\ref{fig:Bp}a, we show the $g$ dependence of $\braket{U_p}$, with different lattice volume $M_x\times M_y$ and a given Fock state cutoff $\Lambda=1$. The points are our results, the black dotted lines are the analytic weak and strong coupling expansion results in \cite{Horsley:1981gj} and \cite{Balian:1974xw}, and the grey region is the bootstrap allowed region of $\braket{U_p}$ that is presented in \cite{Li:2024wrd}. We find that our result agrees with those results qualitatively, but there are also some minor deviations between them. This is because our result is calculated by non-compact QED, while their result is obtained by compact QED. We also test the Fock state cutoff $\Lambda$ dependence of $\braket{U_p}$ in Fig.~\ref{fig:Bp}b and Fig.~\ref{fig:Bp}c, with a given lattice volume $M_x=M_y=10$. It is shown in Fig.~\ref{fig:Bp}c that the precision of $\Lambda = 2$ result, which only uses $3$ states per lattice site, can reach more than $10^{-6}$ in $g^2 a<1$ region.

    \begin{figure}
        \centering
        \includegraphics[width=0.98\linewidth]{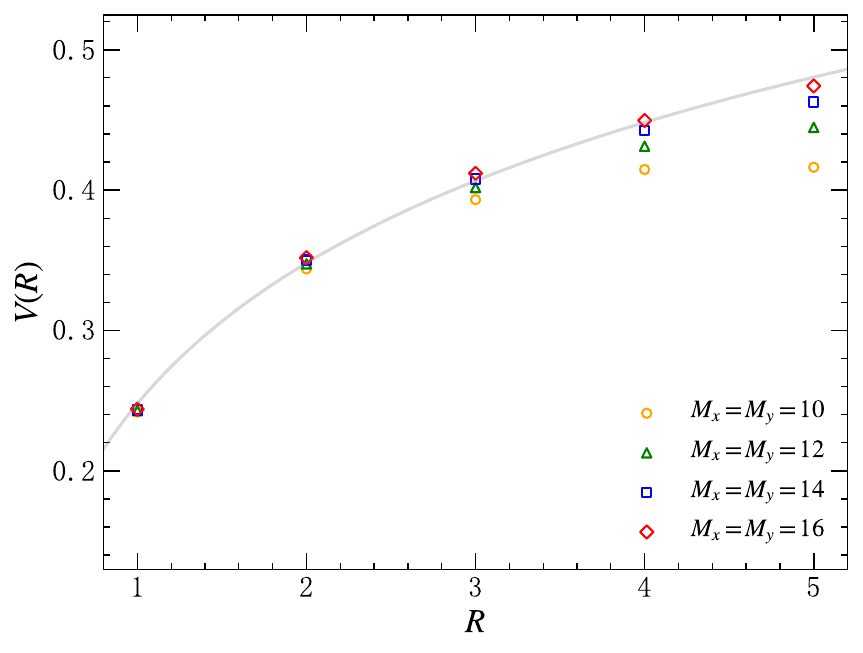}
        \caption{The static potential extracted from Eq.~(\ref{eq:stpot}) in different lattice volume $M_x\times M_y$, with fixing $g=0.7$, $a=1$ and $\Lambda=3$. The point data comes from the lattice calculation, while the grey line comes from fitting the $M_x=M_y=16$ data using $V(R)=a_V\log R+b_V$. The fitting result is $a_V=0.1445$ and $b_V=0.2479$.}
        \label{fig:VR}
    \end{figure}
    
\begin{figure}
        \centering
        \includegraphics[width=0.98\linewidth]{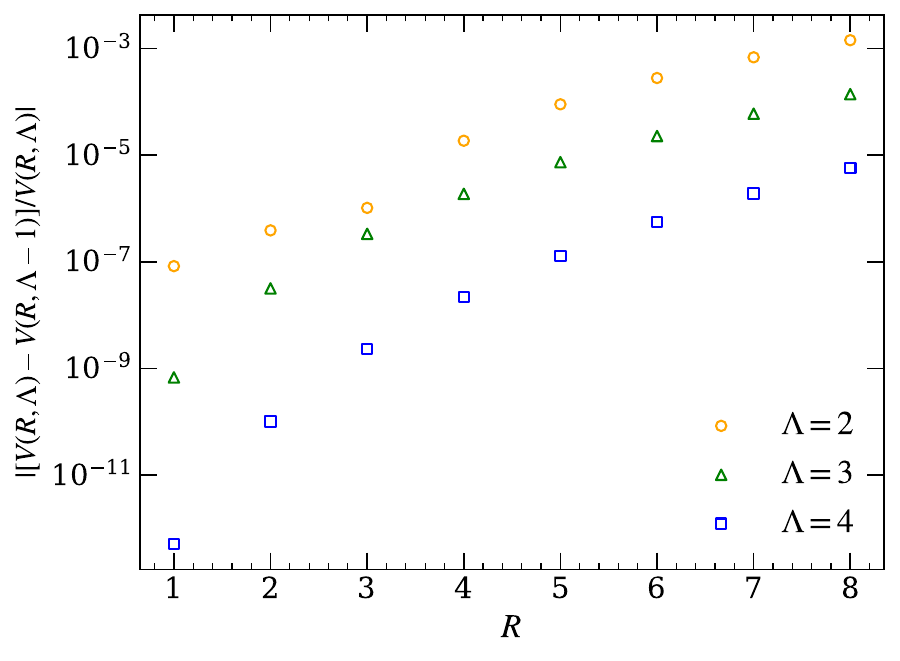}
        \caption{The Fock state truncation $\Lambda$ dependence of static potential, with fixing $M_x=M_y=16$, $g=0.7$, and $a=1$.}
        \label{fig:wl_Lam}
    \end{figure}

    We also calculated the long-time behavior of the Wilson loop. The Wilson loop in the $(2+1)$-dimensional CG QED can be written as:
    \begin{align}\label{eq:wloop}
        W(R,t) =& \left[e^{iga_t\hat{A}_0(\mathbf{0})}e^{i\hat{H}a_t}\right]^{M_t}\left\{\prod_{s=1}^{R}U^\dagger_i\left((R-s)\hat{i}\right)\right\}\nonumber \\
        &\times \left[e^{-i\hat{H}a_t}e^{-iga_t\hat{A}_0(R\hat{i})}\right]^{M_t}\left[\prod_{s=0}^{R-1}U_i(s \hat{i})\right]\,.
    \end{align}
    $i=1,2$ will give the same result because of the rotational symmetry of the CG. Considering a pair of static positron and electron that stay on the positions $\mathbf{n_+}$ and $\mathbf{n_-}$, then the gauge potential $\hat{A}_0(\mathbf{n})$ is given by
    \begin{align}
        \hat{A_0}(\mathbf{n};\mathbf{n}_+,\mathbf{n}_-) = \sum_{\mathbf{p}\not=0} \frac{g}{\hat{E}^2_{\mathbf{p}}} \left(e^{-i\mathbf{p}\cdot(\mathbf{n}-\mathbf{n}_+)}-e^{-i\mathbf{p}\cdot(\mathbf{n}-\mathbf{n}_-)}\right)\,.
    \end{align}  
    Then, we can obtain the Wilson loop related to a pair of static electric positron sources by placing $\hat{A}_0(\mathbf{0})=\hat{A}_0(\mathbf{0};R\hat{i},\mathbf{0})$ and $\hat{A}_0(R\hat{i})=\hat{A}_0(R\hat{i};R\hat{i},\mathbf{0})$ in the Eq.~(\ref{eq:wloop}). According to potential NRQCD \cite{Brambilla:1999xf}, the static potential 
    $V(r)$ can be extracted by calculating the logarithm of the VEV of the Wilson loop $\braket{W(R,t)}$:
    \bea\label{eq:stpot}
        \frac{1}{-it}\ln\braket{W(R,t)} = V(R)+O\left(\frac{1}{t}\right)\,.
    \eea
    It is well known that the electron-positron pair has a logarithmic potential in (2+1)-dimensional non-compact QED. So for a given $R$, the real part of $\frac{1}{-it}\ln\braket{W(R,t)}$ will converge to a non-zero real number, and the imaginary part will tend to zero in the long-time limit. That is what we have seen in Fig.~\ref{fig:wloop}. In Fig.~\ref{fig:VR}, we show the static potential extracted from the long-time Wilson loop in different lattice volumes. We find that our result can be fitted by a logarithmic function, which is consistent with the potential in (2+1)-dimensional QED. Finally, we test the $\Lambda$ dependence of $V(R,\Lambda)$ in Fig.~\ref{fig:wl_Lam}. We find that when $\Lambda=3$, the relative error of $V(R<5)$ is less than $10^{-5}$, which shows the rationality of choosing $\Lambda=3$ in Fig.~\ref{fig:VR}.

\section{Summary and outlook}\label{sec:s_a_o}
    In this study, based on the CG, we propose a new formalism for simulating LGT on a quantum computer, encompassing the discretization of fields and the mapping of these fields to qubits. As a preliminary exploration, we discuss the quantum simulations of QED. We discretize the gauge fields directly in our formalism. This formalism preserves Gauss's law and the CG condition by solving the lattice photon polarization vectors either numerically or analytically. One significant advantage of this approach is its efficiency in simulating hadron scattering processes on future fault-tolerant quantum computers. We also discuss the complexity of our formalism, including the qubit cost of encoding fields and the gate cost of simulating time evolution. We also investigate the convergence behavior in the limit $\Lambda\to \infty$ of our formalism. Compared with K-S formalism, our formalism costs more gates to simulate time evolution. Finally, the results for $\braket{U_p}$, the long-time behavior of the Wilson loop, and the static potential $V(R)$ validate the rationality of our formalism.

    Extending this formalism to non-Abelian cases presents another intriguing area of investigation. The Hamiltonian for lattice SU(N) gauge theory in the CG has been discussed in \cite{Baaquie:1985py}. However, unlike the Abelian case, the CG in non-Abelian theories cannot eliminate all residual degrees of freedom due to the presence of Gribov ambiguity \cite{Gribov:1977wm}. Consequently, addressing Gribov ambiguity on quantum computers constitutes a primary challenge for extending our formalism to non-Abelian gauge theories. Another challenge involves solving Gauss's law for non-Abelian gauge theories in the CG framework. Strategies for tackling these two challenges will be explored in our future work.

\begin{acknowledgments}
We thank Xingyu Guo, Wai Kin Lai, Lingxiao Wang, Hongxi Xing, Dan-Bo Zhang, and Kai Zhou for helpful discussions. This research is supported by Guangdong Major Project of Basic and Applied Basic Research with No. 22020B0301030008, 2022A1515010683, and by the National Natural Science Foundation of China with Project No. 12035007, 12022512.
\end{acknowledgments}

\normalem
\bibliographystyle{h-physrev5}   
\bibliography{ref.bib}

\clearpage
\onecolumngrid
\section*{Supplemental Material}

% 重定义编号方式
\setcounter{section}{0}
\renewcommand{\thesection}{S\arabic{section}}

\setcounter{equation}{0}
\renewcommand{\theequation}{S\arabic{equation}}

\setcounter{figure}{0}
\renewcommand{\thefigure}{S\arabic{figure}}

\setcounter{table}{0}
\renewcommand{\thetable}{S\arabic{table}}

\section{QED Hamiltonian in Coulomb gauge}
In the CG, the QED Hamiltonian is given by \cite{Weinberg:1995mt}
\bea\label{eq:conham}
    H&=\int d^3 x\left[\frac{1}{2} \mathbf{\Pi}_{\perp}^2+\frac{1}{2}(\nabla \times \mathbf{A})^2\right.\nn
    &\quad\left.-\mathbf{J} \cdot \mathbf{A}+\frac{1}{2} J^0 A^0\right]+H_{\mathrm{M}}\nn
    &\equiv H_E+H_B+H_{I}+H_{V}+H_M\,,
\eea
with four constraints
\bea\label{eq:con}
    &\chi_1(\mathbf{x})\equiv\Pi^0 = 0\,,\nn
    &\chi_2(\mathbf{x})\equiv\partial_i \Pi^i - J^0 = 0\,,\nn
    &\chi_3(\mathbf{x})\equiv\partial_i A^i = 0\,,\nn
    &\chi_4(\mathbf{x})\equiv\partial_i \Pi^i - \partial_i \partial^i A^0 = 0\,,
\eea
where $\Pi^i_{\perp} = {\Pi}^i(x)-\partial^i A^0$ is the conjugate momentum to the gauge field $A_i$, $J^\mu = g\bar{\psi} \gamma^\mu \psi$ is the electric current and $H_M$ is the Hamiltonian of matter field
\bea
    H_M = \int d^3x \bar{\psi} (-i \gamma^i\partial_i +m)\psi\,.
\eea
Note that all of the residual gauge freedoms are eliminated in the CG. This can be explained as follows: suppose $\partial_i A^i(x)=0$ and there is another potential $A^{\prime \mu}$ that also satisfies the CG. Then We have $\partial_i A^{\prime i}(x) = \partial_i\partial^i \eta(x)=0$ for all times, where $\eta$ is a scalar function. By choosing boundary conditions such that $\eta(x)$ vanishes at infinity, it follows that $\eta(x) = 0$. Therefore, there are no nontrivial gauge transformations that can simultaneously preserve both the Hamiltonian and the CG condition. Hence, selecting the CG can satisfy requirement 5) in the introduction. This is not the case for the temporal gauge, where it is possible to maintain $A^0(x) = 0$ by transforming $A^\mu$ with an arbitrary time-independent scalar function $\eta(\mathbf{x})$. In the CG, Gauss's law is a constraint for $A^0$. After solving Gauss's law, $A^0$ can be written as
\bea
    A^0(t,\mathbf{x})=\int d^3 y \frac{J^0(t,\mathbf{y})}{4 \pi|\mathbf{x}-\mathbf{y}|}\,.
\eea
To preserve Gauss's law and CG condition for all the time, $A^i$ and their conjugate momenta should satisfy the Dirac bracket quantization relations
\bea\label{eq:qua}
    & {\left[A_i(\mathbf{x}), \Pi^j_\perp(\mathbf{y})\right]=i \delta_i^j \delta^3(\mathbf{x}-\mathbf{y})+i \frac{\partial^2}{\partial x^j \partial x^i}\left(\frac{1}{4 \pi|\mathbf{x}-\mathbf{y}|}\right),} \nn
    & {\left[A_i(\mathbf{x}), A_j(\mathbf{y})\right]=\left[\Pi^i_\perp(\mathbf{x}), \Pi^j_\perp(\mathbf{y})\right]=0}\,.
\eea
The fermion fields satisfy the anti-commutation relations
\bea
    &\{\psi_\alpha(\mathbf{x},\psi^\dagger_\beta(\mathbf{y}))\} = \delta(\mathbf{x}-\mathbf{y})\,,\nn
    &\{\psi_\alpha(\mathbf{x}),\psi_\beta(\mathbf{y})\} = \{\psi^\dagger_\alpha(\mathbf{x}),\psi^\dagger_\beta(\mathbf{y})\} = 0
\eea
In the Schrodinger picture, the $A_\mu(\mathbf{x})$ and $\Pi_{\perp \mu}(\mathbf{x})$ can be expanded as
\bea\label{eq:expan}
    &A_i (\mathbf{x}) = \int \frac{d^3p}{(2\pi)^3\sqrt{2\abs{\mathbf{p}}}} \sum_{r} \left[\epsilon^r_i (\mathbf{p}) a^r_\mathbf{p} e^{i\mathbf{p}\cdot \mathbf{x}}+{\rm H.c.}\right]\,,\nn
    &\Pi_{\perp i} (\mathbf{x}) = \int \frac{d^3p}{(2\pi)^3}\sqrt{\frac{\abs{\mathbf{p}}}{2}} \sum_{r} \left[-i\epsilon^r_i (\mathbf{p}) a^r_\mathbf{p} e^{i\mathbf{p}\cdot \mathbf{x}}+{\rm H.c.}\right]\,,
\eea
where $\epsilon^r_\mu (\mathbf{p}),\, r = \pm 1$ are the polarization vectors, which can be written as $\epsilon^r(\mathbf{p}) = \mathcal{R}(\mathbf{p}) \epsilon^r$. $\mathcal{R}(\mathbf{p})$ is the standard rotation connect the direction of $z$-axis and $\mathbf{p}$. $\epsilon^{\pm}_\mu = (0,1/\sqrt{2},\pm i/\sqrt{2},0)$ are the standard polarization vectors of the photon. To be consistent with the commutation relations in Eq.~(\ref{eq:qua}), the commutation relations between $a^{r\dagger}_{\mathbf{p}}$ and $a^{r\prime}_{\mathbf{p}^\prime}$ should be
\bea\label{eq:momquant}
    &[a^{r}_{\mathbf{p}^\prime},a^{s\dagger}_{\mathbf{p}}] = \delta_{rs}\delta_{\mathbf{p}'\mathbf{p}}\,,\nn
    &[a^{r}_{\mathbf{p}^\prime},a^{s}_{\mathbf{p}}] = [a^{r\dagger}_{\mathbf{p}^\prime},a^{s\dagger}_{\mathbf{p}}] = 0\,.
\eea
It is not hard to check that the mode expansion in Eq.~(\ref{eq:expan}) satisfies the quantization condition Eq.~(\ref{eq:qua}) and the CG condition.

\section{Discretizations of Coulomb gauge QED}
    Both gauge fields and fermion fields need to be discretized. The discretization of the gauge field will be discussed first. Because there are no residual gauge degrees of freedom in the CG, we can discrete the gauge fields $A_i(\mathbf{x})$ and their conjugate momentum $\Pi^i(\mathbf{x})$ directly
    \bea
        &\hat\partial^R_i \hat A^j(\mathbf{n}) \equiv \frac{1}{a} [\hat A^j(\mathbf{n}+a\hat{i})-\hat A^j(\mathbf{n})]\,,\nn
        &\hat\partial^R_i \hat \Pi^j(\mathbf{n}) = \frac{1}{a} [\hat{\Pi}^j(\mathbf{n}+a\hat{i})-\hat{\Pi}^j(\mathbf{n})]\,,
    \eea
    where $\hat{i}$ is the unit vector along the $i$ direction, $a$ is the lattice spacing, $\hat A$ is the discrete version of gauge field $A$, and the space point can be written as $\mathbf{n} = (n_x,n_y,n_z),\, n_x,n_y,n_z = 0,1...,M-1$, so there are $M^d$ lattice sites in $d$-dimensional space. The Laplace operator $\Delta = -\partial_i\partial^i$ appears in the Hamiltonian of CG QED. The discretized version of Laplace operator $\hat{\Delta}$ is
    \bea
        \hat{\Delta}\hat A_j(\mathbf{n}) \equiv \frac{1}{a^2}\sum_i [\hat A_j(\mathbf{n}+a\hat{i})-2\hat A_j(\mathbf{n})+\hat A_j(\mathbf{n}-a\hat{i})]\,.
    \eea
    
    For a constrained system, the constraints in Eq.~(\ref{eq:con}) should be modified to the lattice version
    \bea\label{eq:latcon}
        &\hat{\chi}_1(\mathbf{n})\equiv\hat{\Pi}^0 = 0\,,\nn
        &\hat{\chi}_2(\mathbf{n})\equiv\hat\partial^R_i \hat{\Pi}^i - J^0 = 0\,,\nn
        &\hat{\chi}_3(\mathbf{n})\equiv\hat\partial^R_i \hat{A}^i = 0\,,\nn
        &\hat{\chi}_4(\mathbf{n})\equiv\hat\partial^R_i \hat{\Pi}^i + \hat{\Delta} \hat{A}^0 = 0\,.
    \eea
    The $\hat A^0$ can be obtained by solving $\hat{\Delta}\hat{A}^0 = -J^0$
    \bea\label{eq:latA0}
        \hat{A}^0(\mathbf{n}) = \sum_{\mathbf{m}} \sum_{\mathbf{p}\not=0} \frac{J^0(\mathbf{m})}{M^3\hat{E}_{\mathbf{p}}^2}e^{-i\mathbf{p}\cdot(\mathbf{n}-\mathbf{m})}\,.
    \eea
    We sum over the non-vanishing momentum mode because the photon with $\mathbf{p} = 0$ is non-physical. The $\hat{E}_{\mathbf{p}}$ is the lattice dispersion relation of photons, which can be obtained by plugging the plane wave solutions into the equation of motion $[(\partial_0)^2-\hat{\Delta}]\hat A_j = 0$,
    \bea\label{eq:latEq}
        \hat{E}_{\mathbf{p}} \equiv \sqrt{\frac{4}{a^2}\sum_{i} \sin^2 \left(\frac{p^i a}{2}\right)}\,.
    \eea
    Eq.~(\ref{eq:latA0}) demonstrates that Gauss's law with dynamical fermions in QED can be naturally resolved on the lattice, thereby meeting requirement 2) outlined in the introduction. According to the Dirac bracket quantization, the commutation relations in Eq.~(\ref{eq:qua}) should also be modified because the constraints change to its lattice version in Eq.~(\ref{eq:latcon})
    \bea\label{eq:latqua}
        &\left[\hat{A}_i(\mathbf{n}), \hat{\Pi}^j(\mathbf{m})\right]=i \delta_j^i \delta_{\mathbf{n},\mathbf{m}}+\frac{i\sum_{\mathbf{p}}(e^{-i p^i a}-1)(e^{i p^j a}-1)}{M^3 a^2 \hat{E}_{\mathbf{p}}^2}\nn
        &\left[\hat{A}_i(\mathbf{n}), \hat{A}_j(\mathbf{m})\right]=\left[\hat{\Pi}^i(\mathbf{n}), \hat{\Pi}^j(\mathbf{m})\right]=0\,.
    \eea
    Now, the discrete version of Eq.~(\ref{eq:expan}) can be written as
    \bea\label{eq:disc_SM}
        &\hat{A}_i (\mathbf{n}) = \sum_{\mathbf{p}\not= \mathbf{0}} \frac{1}{\sqrt{2\hat{E}_{\mathbf{p}} M^3}} \sum_{r} \left[\hat{\epsilon}^r_i (\mathbf{p}) a^r_\mathbf{p} e^{i\mathbf{p}\cdot \mathbf{n}}+{\rm H.c.}\right]\,,\nn
        &\hat{\Pi}_{\perp i} (\mathbf{n}) = \sum_{\mathbf{p}\not = \mathbf{0}} \sqrt{\frac{\hat{E}_{\mathbf{p}}}{2M^3}} \sum_{r} \left[-i\hat{\epsilon}^r_i (\mathbf{p}) a^r_\mathbf{p} e^{i\mathbf{p}\cdot \mathbf{n}}+{\rm H.c.}\right]\,,
    \eea
    where $\hat{\epsilon}^r_i$ are the lattice polarization vectors, which need to be determined later. Eq.~(\ref{eq:disc_SM}) shows that $\hat{A}_i(\mathbf{n})$ are Hermitian operators, enabling us to express unitary Wilsons line as $U_i(\mathbf{n}) = \exp(-iga\hat{A}_i(\mathbf{n}))$, thereby fulfilling requirement 3) outlined in the introduction. On the lattice of momentum space, the momentum $\mathbf{p}$ can be evaluated as $\mathbf{p} = (2\pi k_x/Ma,2\pi k_y/Ma,2\pi k_z/Ma)$, with $k_x,k_y,k_z=0,\pm1,...,\pm \lfloor M/2 \rfloor$. To preserve the lattice CG condition, Gauss's law and the lattice Dirac bracket quantization conditions in Eq.~(\ref{eq:latqua}), the lattice polarization vectors $\hat{\epsilon}^r_i$ should satisfy
    \bea\label{eq:lateps_SM}
        &\sum_i (e^{i p^i a}-1) \hat{\epsilon}^r_i(\mathbf{p}) = 0\,, \nn
        &\sum_i \hat{\epsilon}^r_i(\mathbf{p}) \hat{\epsilon}^{s}_i(\mathbf{p}) = \delta_{rs}\,,\nn
        &\sum_{r} \hat{\epsilon}^r_i(\mathbf{p}) \hat{\epsilon}^r_j(\mathbf{p}) = \delta_{ij} -\frac{1}{\hat{E}_{\mathbf{p}}^2} (e^{-i p^i a}-1) (e^{i p^j a}-1)\,.
    \eea
    So, $\hat{\epsilon}^r_i$ can be obtained by solving Eq.~(\ref{eq:lateps_SM}) numerically or analytically.
    
    It's worth noting that both the solution of $\hat{A}_0$ in Eq.~(\ref{eq:latA0}) and the solution of $\hat{\epsilon}^r_i(\mathbf{p})$ in Eq.~(\ref{eq:lateps_SM}) do not depend on the truncation of photon Fock states. This means four constraints in Eq.~(\ref{eq:latcon}) will be preserved accurately after the digitization of gauge fields, which is important for simulating gauge field theories on quantum computers.
    
    For the fermion fields, they can be discretized by various methods, here, we use Wilson fermion. Fermion fields can be discredited naively in this discretization scheme, but one needs to add a Wilson term $\hat{H}_W$ into the Hamiltonian. After doing so, the $\hat{H}_E+\hat{H}_B$, $\hat H_I$, $\hat H_V$, $\hat H_M$ and $\hat H_W$ can be written as (we set $a=1$ here)
    \bea\label{eq:dis_ham_SM}
        &\hat{H}_E + \hat{H}_B = \sum_{\mathbf{p}\not = \mathbf{0}} \sum_{r} \hat{E}_{\mathbf{p}} a^{r\dagger}_{\mathbf{p}} a^{r}_{\mathbf{p}}\,,\nn
        &\hat{H}_I = \sum_{\mathbf{n},i} \sum_{\mathbf{p}\not=0} \sum_{r} \frac{J^i(\mathbf{n})}{M^{\frac{3}{2}}\sqrt{2\hat{E}_{\mathbf{p}}}}\left[\hat{\epsilon}^r_i (\mathbf{p}) a^r_\mathbf{p} e^{i\mathbf{p}\cdot \mathbf{n}}+{\rm H.c.}\right]\,,\nn
        &\hat{H}_V = \frac{1}{2}\sum_{\mathbf{n},\mathbf{m}} \sum_{\mathbf{p}\not=0} \frac{J^0(\mathbf{n})J^0(\mathbf{m})}{M^3 \hat{E}_{\mathbf{p}}^2}e^{-i\mathbf{p}\cdot(\mathbf{n}-\mathbf{m})}\,,\nn
        &\hat{H}_M = \sum_{\mathbf{n},i} \bar{\psi}(\mathbf{n})\left[-i \gamma^i \frac{\psi(\mathbf{n}+\hat{i})-\psi(\mathbf{n}-\hat{i})}{2}+m\bar{\psi}\psi\right]\,,\nn
        &\hat{H}_W = \sum_{\mathbf{n}}-\frac{w}{2} \bar{\psi}(\mathbf{n}) \hat{\Delta} \psi(\mathbf{n})\,,
    \eea
    where $0<w<1$ is the coefficient of the Wilson term. Finally, the discretized CG QED Hamiltonian is $\hat{H} = \hat{H}_E+\hat{H}_B+\hat{H}_I+\hat{H}_V+\hat{H}_M+\hat{H}_W$.

\section{Estimation of the truncation error}
In this section, we will estimate the error associated with truncating the photon Fock state.
\subsection{A toy model}
We first consider a toy model
\bea
    H_{\rm toy} = \omega(a^\dagger a) + g(a+a^\dagger)\,,
\eea
with the commutation relation $[a,a^\dagger] = 1$. The eigenstate of occupation operator $\hat{\mathcal{N}}=a^\dagger a$ needs to be truncated: $a^\dagger\ket{\mathcal{N}=\Lambda-1}=0$. Now, the problem is: if there is a state $\ket{\psi}$ and the vacuum state $\ket{\Omega}$, which has $E = \bra{\psi}H\ket{\psi} - \braket{\Omega|H|\Omega}$, and we want the error of the truncated state $\ket{\psi(\Lambda)}$ below $\varepsilon$. Then, how can we choose the truncation $\Lambda$? This is equivalent to letting the probability $P\left({\hat{\mathcal{N}}}>{\Lambda-1}\right)$ smaller than $\varepsilon$ when measure the operator $\hat{\mathcal{N}}$ in the quantum state $\psi$. To solve this problem, we first define a new creation and annihilation operator $\bar{a}$ and $\bar{a}^\dagger$, then calculate the $\braket{\Omega|H|\Omega}$:
\bea
    \bar{a} = a+\frac{g}{\omega}\,, \ \bar{a}^\dagger = a^\dagger+\frac{g}{\omega}\,.
\eea
The new creation and annihilation operators also satisfy the commutation relation $[\bar{a},\bar{a}^\dagger]=1$. Using those operators, the Hamiltonian can be rewritten as
\bea
    H_{\rm toy} = \omega \bar{a}^\dagger \bar{a} - \frac{g^2}{\omega}.
\eea
From the above equation, we can solve the vacuum energy of $H_{\rm toy}$:
\begin{align}
    \braket{\Omega|H|\Omega} = E_0 = -\frac{g^2}{\omega}\,.
\end{align}
We then have $\braket{\psi|H_{\rm toy}|\psi} = E-g^2/\omega$.

\begin{figure}[htbp]
            \centering
            \includegraphics[width=0.98\linewidth]{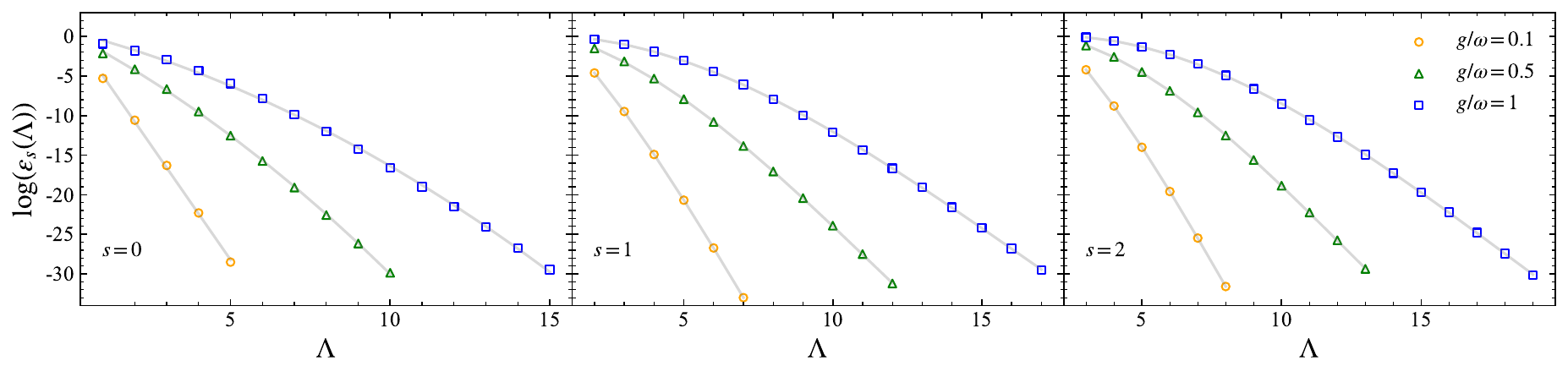}
            \caption{$\log(\varepsilon_s)$ dependence of $\Lambda$ from weak to strong coupling cases. $s=0,1,2$ represent vacuum, first and second excited states of $H_{\rm toy}$ respectively. All of the data in this figure can be fitted by grey solid lines, which are parameterized as $-a\Lambda^2-b\Lambda-c$.}
            \label{fig:state_err}
    \end{figure}

Next, we need to estimate the upper bound of the expectation value $|\braket{\psi|a^\dagger + a|\psi}|$. To achieve this, we need to define the operators $X$ and $P$:
\begin{align}
    X = \frac{a+a^\dagger}{\sqrt{2}}\,, \ P = \frac{a-a^\dagger}{\sqrt{2}i}\,.
\end{align}
From the above equations, we can find that
\begin{align}
    X^2 = 2\hat{\mathcal{N}}-P^2+1\,.
\end{align}
Then we have the following inequality:
\begin{align}\label{eq:ineaad}
    |\braket{a+a^\dagger}| \leq \sqrt{\braket{(a+a^\dagger)^2}} = \sqrt{\braket{2X^2}}=\sqrt{2\braket{2\hat{\mathcal{N}}-P^2+1}} \leq \sqrt{\braket{4\hat{\mathcal{N}}+2}}\,,
\end{align}
where $\braket{...}$ represents the expectation value $\braket{\psi|...|\psi}$. From the $\braket{H_{\rm toy}} = E-g^2/\omega$ and the Eq.~\eqref{eq:ineaad}, we can obtain:
\begin{align}
    E - g^2/\omega \geq \omega\braket{\hat{\mathcal{N}}} - |g\braket{a+a^\dagger}| \geq \omega\braket{\hat{\mathcal{N}}} - |g| \sqrt{4\braket{\hat{\mathcal{N}}}+2}.
\end{align}
Then, we can solve the upper bound of the $\braket{\hat{\mathcal{N}}}$ by solving the above inequation:
\begin{align}
    \braket{\hat{\mathcal{N}}} \leq \left(\frac{|g|}{\omega} + \sqrt{\frac{E}{\omega}+\frac{1}{2}}\right)^2-\frac{1}{2}\,.
\end{align}
The above equation gives the correct limit in $E = 0$ and $g = 0$. Once known the $\braket{\hat{\mathcal{N}}}$, we can obtain the probability $P(\hat{\mathcal{N}}>\Lambda) = \sum_{\mathcal{N}=\Lambda}^{\infty} |\braket{\psi|\mathcal{N}}|^2$ by using the Markov's inequality:
\begin{align}\label{eq:MKine}
    P(Y\geq \Lambda) \leq \frac{\braket{Y}}{\Lambda}\,,
\end{align}
where $Y$ is an arbitrary not-negative random variable. In our case, we have $Y = \hat{\mathcal{N}}$, so:
\begin{align}
    P(\hat{\mathcal{N}} \geq \Lambda) \leq \frac{\braket{\hat{\mathcal{N}}}}{\Lambda} \leq \frac{1}{\Lambda}\left[\left(\frac{|g|}{\omega} + \sqrt{\frac{E}{\omega}+\frac{1}{2}}\right)^2-\frac{1}{2}\right] = \varepsilon_s\,.
\end{align}
It can be seen that when $\varepsilon_s \to 0$, the amplitude square $\sum_{\mathcal{N}=\Lambda}^{\infty} |\braket{\psi|\mathcal{N}}|^2$ will tend to zero, then the truncated tends to the untruncated state. The final equal sign in the above equation gives us the bound of truncation $\Lambda$ with given precision $\varepsilon_s$:
\begin{align}\label{eq:Lam_eps}
    \Lambda = \frac{1}{\varepsilon_s} \left[\left(\frac{|g|}{\omega} + \sqrt{\frac{E}{\omega}+\frac{1}{2}}\right)^2-\frac{1}{2}\right]\,.
\end{align}
We find that as $\varepsilon_s \propto 1/\Lambda$ in Eq.~\eqref{eq:Lam_eps}. However, this is the worst case because we consider the most general state $\ket{\psi}$ when we deduce Eq.~\eqref{eq:Lam_eps}. As shown in Fig.~\ref{fig:state_err}, if we consider the ground and lower excited state of $H_{\rm toy}$, $\varepsilon_s$ converges exponentially to $0$ as $\Lambda$ increases.

\subsection{QED in CG}
Although the toy model discussed in the last subsection is simple, most of the derivation in the toy model can be generalized to CG QED. Suppose the physics we are interested in has energy scale $E = \braket{\Psi|\hat{H}|\Psi} \equiv \braket{\hat{H}}$, then we have:
\begin{align}\label{eq:HEBIqs}
    \braket{\hat{H}_{EB}(\mathbf{q},s)+\hat{H}_I(\mathbf{q},s)}= E - \braket{\hat{H}_V} - \braket{\hat{H}_{M}+\hat{H}_W} - \sum_{\mathbf{p}\not=\mathbf{q}\, {\rm and}\,r\not=s} \braket{\hat{H}_{EB}(\mathbf{p},r)+\hat{H}_I(\mathbf{p},r)}\,,
\end{align}
where
\begin{align}
    &\hat{H}_{EB}(\mathbf{p},r) \equiv \hat{E}_{\mathbf{p}} a^{r\dagger}_{\mathbf{p}} a^r_{\mathbf{p}}\,, \nonumber\\
    &\hat{H}_I(\mathbf{p},r) \equiv \sum_{\mathbf{n},i} \frac{J^i(\mathbf{n})}{M^{\frac{d}{2}}\sqrt{2 \hat{E}_{\mathbf{p}}}} \left[\hat{\epsilon}^r_i(\mathbf{p}) a^r_{\mathbf{p}} e^{i \mathbf{p} \cdot \mathbf{n} + {\rm H.c.}}\right]\,.
\end{align}
According to the discussions of the toy model, we need to estimate the upper bound of $|\braket{\hat{H}_I(\mathbf{q},s)}|$ to obtain the upper bound of the $\braket{\hat{\mathcal{N}}_{\mathbf{q},s}}$. To achieve this, we need to calculate the lower bound of $\braket{\hat{H}_V}$, $\braket{\hat{H}_M + \hat{H}_W}$, and every $\braket{\hat{H}_{EB}(\mathbf{p},r)+\hat{H}_I(\mathbf{p},r)}$.

Firstly, we estimate the lower bound of $\hat{H}_V$. $J^0(\mathbf{n})$ is the only operator appearing in $\hat{H}_V$. At each space point, the charge density must be equal to or larger than $g$, so the operator $J^0(\mathbf{n})+g$ must be positive definite:
\begin{align}
     \frac{1}{2}\sum_{\mathbf{n},\mathbf{m}} \sum_{\mathbf{p}\not=0} \frac{\braket{[J^0(\mathbf{n})+g][J^0(\mathbf{m})+g]}}{M^d \hat{E}_{\mathbf{p}}^2}e^{-i\mathbf{p}\cdot(\mathbf{n}-\mathbf{m})} = \braket{\hat{H}_V} + \sum_{\mathbf{n},\mathbf{m}} \sum_{\mathbf{p}\not=0} \frac{g \braket{J^0(\mathbf{n})}}{M^d \hat{E}^2_{\mathbf{p}}} e^{-i\mathbf{p}\cdot(\mathbf{n}-\mathbf{m})} + \frac{g^2}{2} \sum_{\mathbf{n},\mathbf{m}} \sum_{\mathbf{p}\not=0} \frac{e^{-i\mathbf{p}\cdot(\mathbf{n}-\mathbf{m})}}{M^d \hat{E}^2_{\mathbf{p}}} \geq 0\,.
\end{align}
Then, we have
\begin{align}\label{eq:lobHV}
    \braket{\hat{H}_V} &\geq - \sum_{\mathbf{n},\mathbf{m}} \sum_{\mathbf{p}\not=0} \frac{g \left|\braket{J^0(\mathbf{n})}\right|}{M^d \hat{E}^2_{\mathbf{p}}} \left|e^{-i\mathbf{p}\cdot(\mathbf{n}-\mathbf{m})}\right| - \frac{g^2}{2} \sum_{\mathbf{n},\mathbf{m}} \sum_{\mathbf{p}\not=0} \frac{\left|e^{-i\mathbf{p}\cdot(\mathbf{n}-\mathbf{m})}\right|}{M^d \hat{E}^2_{\mathbf{p}}}\nonumber \\
    &\geq - \sum_{\mathbf{n},\mathbf{m}} \sum_{\mathbf{p}\not=0} \frac{g \left|\braket{J^0(\mathbf{n})}\right|}{M^d \hat{E}^2_{\mathbf{p}}} - \frac{g^2}{2} \sum_{\mathbf{n},\mathbf{m}} \sum_{\mathbf{p}\not=0} \frac{1}{M^d \hat{E}^2_{\mathbf{p}}} \nonumber\\
    &\geq -\sum_{\mathbf{p}\not=0} \frac{g^2 M^d}{\hat{E}^2_{\mathbf{p}}} - \sum_{\mathbf{p}\not=0}\frac{g^2 M^d}{2\hat{E}^2_{\mathbf{p}}} \nonumber \\
    &\geq -\sum_{\mathbf{p}\not=0} \frac{3 g^2 M^{d+2}}{8\pi^2} \approx -\frac{3g^2 M^{2d+2}}{8\pi^2}\,.
\end{align}
In the second line of the above inequality, we use $|e^{-i\mathbf{p}\cdot(\mathbf{n}-\mathbf{m})}| \leq 1$. The third line is because $\braket{J^0(\mathbf{n})} \leq g$. The forth line is because $\hat{E}_\mathbf{p} = 2\sqrt{\sum_i \sin^2(2\pi k_i/2M)} \geq 2\sqrt{ \sin^2(\pi/M)} \approx 2\pi/M$.

The upper bound of $\braket{\hat{H}_M+\hat{H}_W}$ is given in \cite{Yao:2025uxz}, which is
\begin{align}\label{eq:lobHMW}
    \braket{\hat{H}_M+\hat{H}_W} \geq -2 M^d \sqrt{3+m^2+12mw+36w^2}\,.
\end{align}

The upper bound of every $\braket{\hat{H}_{EB}(\mathbf{p},r)+\hat{H}_I(\mathbf{p},r)}$ also needs to be estimated. As we have done in the toy model case, we need to define a new annihilation operator $\bar{a}^{r,i}_{\mathbf{p}}$:
\begin{align}
    \bar{a}^{r,i}_{\mathbf{p}} = a^r_{\mathbf{p}} + \sum_{\mathbf{n}} \frac{dJ^i(\mathbf{n})}{\sqrt{2}M^{\frac{d}{2}}\hat{E}_{\mathbf{p}}^{\frac{3}{2}}}\hat{\epsilon}^{r*}_i(\mathbf{p}) e^{-i\mathbf{p}\cdot \mathbf{m}}\,.
\end{align}
It is worth noting that the index $i$ has not been contracted in the definition of $\bar{a}^{r,i}_{\mathbf{p}}$.  It is not so hard to check the operator $\bar{a}^{r,i}_{\mathbf{p}}$ and $\bar{a}^{r,i\dagger}_{\mathbf{p}}$ satisfy:
\begin{align}\label{eq:aad_SM}
    [\bar{a}^{r,i}_{\mathbf{p}},\bar{a}^{s,i\dagger}_{\mathbf{q}}] = [a^r_{\mathbf{p}},a^s_{\mathbf{q}}] = \delta^{rs} \delta_{\mathbf{p},\mathbf{q}}\,.
\end{align}
Using the operator $\bar{a}^{r,i}_{\mathbf{p}}$, the Hamiltonian $\hat{H}_{EB}(\mathbf{p},r)+\hat{H}_I(\mathbf{p},r)$ can be expressed as:
\begin{align}\label{eq:HEBIpr}
    \hat{H}_{EB}(\mathbf{p},r)+\hat{H}_I(\mathbf{p},r) = \sum_i \frac{\hat{E}_{\mathbf{p}}}{d} \bar{a}^{r,i\dagger}_{\mathbf{p}} \bar{a}^{r,i}_{\mathbf{p}} - \sum_i \sum_{\mathbf{n},\mathbf{m}} \frac{dJ^i(\mathbf{n})J^i(\mathbf{m})}{2M^d \hat{E}^2_{\mathbf{p}}} \hat{\epsilon}^r_i(\mathbf{p})\hat{\epsilon}^{r*}_{i}(\mathbf{p}) e^{i\mathbf{p}\cdot(\mathbf{n}-\mathbf{m})}\,.
\end{align}
From Eq.~\eqref{eq:aad_SM}, we know that the first term on the right hand side of Eq.~\eqref{eq:HEBIpr} is positive definite, so that we can find the following lower bound:
\begin{align}\label{eq:lobHEBI}
    \braket{\hat{H}_{EB}(\mathbf{p},r)+\hat{H}_I(\mathbf{p},r)} &\geq - \sum_i \sum_{\mathbf{n},\mathbf{m}} \frac{d\left|\braket{J^i(\mathbf{n})J^i(\mathbf{m})}\right|}{2M^d \hat{E}^2_{\mathbf{p}}} \hat{\epsilon}^r_i(\mathbf{p})\hat{\epsilon}^{r*}_{i}(\mathbf{p}) \left|e^{i\mathbf{p}\cdot(\mathbf{n}-\mathbf{m})}\right| \nonumber \\
    &\geq -\sum_i \sum_{\mathbf{n},\mathbf{m}} \frac{d\left|\braket{J^i(\mathbf{n})J^i(\mathbf{m})}\right|}{2M^d \hat{E}^2_{\mathbf{p}}} \left|e^{i\mathbf{p}\cdot(\mathbf{n}-\mathbf{m})}\right| \nonumber \\
    &\geq -\sum_i \sum_{\mathbf{n},\mathbf{m}} \frac{d\left|\braket{J^i(\mathbf{n})J^i(\mathbf{m})}\right|}{2M^d \hat{E}^2_{\mathbf{p}}}\nonumber\\
    &\geq -\sum_i \sum_{\mathbf{n},\mathbf{m}} \frac{2d g^2}{M^d \hat{E}^2_{\mathbf{p}}} = -\frac{2 d^2 M^d g^2}{\hat{E}^2_{\mathbf{p}}}\,.
\end{align}
The second line of this inequality is due to $\hat{\epsilon}^r_i(\mathbf{p})\hat{\epsilon}^{r*}_{i}(\mathbf{p})\leq 1$, which comes from the normalization of polarization vectors. In the third line, we use the fact that $|e^{i\mathbf{p}\cdot(\mathbf{n}-\mathbf{m})}| \leq 1$. From Eq.~\eqref{eq:JWJi}, we can find that the upper bound of  $|\braket{J^i(\mathbf{n})J^i(\mathbf{m})}|$ is $4g^2$, and then we can obtain the fourth line of the inequality.

Now, the upper bound of $\braket{\hat{H}_{EB}(\mathbf{q},s)+\hat{H}_{I}(\mathbf{q},s)}$ can be obtained by combining the Eq.~\eqref{eq:HEBIqs}, Eq.~\eqref{eq:lobHV}, Eq.~\eqref{eq:lobHMW}, and Eq.~\eqref{eq:lobHEBI}:
\begin{align}
    \braket{\hat{H}_{EB}(\mathbf{q},s)+\hat{H}_{I}(\mathbf{q},s)} &\leq E+\frac{3 g^2 M^{2d+2}}{8\pi^2} + 2M^d \sqrt{3+m^2+12mw+36w^2} + \sum_{\mathbf{p}\not=\mathbf{q}\ {\rm and}\ r\not=s} \frac{2 d^2 g^2 M^d}{\hat{E}^2_{\mathbf{p}}}\nonumber \\
    &\leq E+\frac{3 g^2 M^{2d+2}}{8\pi^2} + 2M^d \sqrt{3+m^2+12mw+36w^2} + \sum_{\mathbf{p}\not=\mathbf{q}\ {\rm and}\ r\not=s} \frac{d^2 g^2 M^{d+2}}{2\pi^2}\nonumber\\
    &< E + \frac{[3+4d^2(d-1)] g^2 M^{2d+2}}{8 \pi^2} + 2M^d  \sqrt{3+m^2+12mw+36w^2}\,.
\end{align}

On the other hand, we have:
\begin{align}
    \braket{\hat{H}_{EB}(\mathbf{q},s)+\hat{H}_{I}(\mathbf{q},s)} = \hat{E}_{\mathbf{q}}\braket{\hat{\mathcal{N}}_{\mathbf{q},s}} + \braket{\hat{H}_{I}(\mathbf{q},s)} \geq \hat{E}_{\mathbf{q}}\braket{\hat{\mathcal{N}}_{\mathbf{q},s}} - |\braket{\hat{H}^S_{I}(\mathbf{q},s)}| - |\braket{\hat{H}^A_{I}(\mathbf{q},s)}|\,,
\end{align}
where $\hat{H_I}(\mathbf{q},s) = \hat{H}^S_I(\mathbf{q},s) + \hat{H}^A_I(\mathbf{q},s)$ and the definition of $\hat{H}^S_I(\mathbf{q},s)$ and $\hat{H}^A_I(\mathbf{q},s)$ are:
\begin{align}
    &\hat{H}^S_I(\mathbf{q},s) = \sum_{\mathbf{n},i} \frac{J^i(\mathbf{n})}{M^{\frac{d}{2}}\sqrt{2\hat{E}_{\mathbf{p}}}} \left[{\rm Re}(\hat{\epsilon}^s_i(\mathbf{p}))\cos(\mathbf{p}\cdot \mathbf{n})-{\rm Im}(\hat{\epsilon}^s_i(\mathbf{p}))\sin(\mathbf{p}\cdot \mathbf{n})\right](a^s_{\mathbf{q}}+a^{s\dagger}_{\mathbf{q}})\,,\nonumber\\
    &\hat{H}^A_I(\mathbf{q},s) = i\sum_{\mathbf{n},i} \frac{J^i(\mathbf{n})}{M^{\frac{d}{2}}\sqrt{2\hat{E}_{\mathbf{p}}}} \left[{\rm Re}(\hat{\epsilon}^s_i(\mathbf{p}))\sin(\mathbf{p}\cdot \mathbf{n})+{\rm Im}(\hat{\epsilon}^s_i(\mathbf{p}))\cos(\mathbf{p}\cdot \mathbf{n})\right](a^s_{\mathbf{q}}-a^{s\dagger}_{\mathbf{q}})\,.
\end{align}
Then, we need to represent the upper bound of $|\braket{\hat{H}^{S(A)}_I(\mathbf{q},s)}|$ as the function of $\braket{\hat{\mathcal{N}}_{\mathbf{q},s}}$. The key point of this step is using the following inequality:
\begin{align}\label{eq:OABine}
    |\braket{\psi|O_A O_B|\psi}| \leq \sqrt{\braket{\psi|O^2_A|\psi}\braket{\psi|O^2_B|\psi}}\,,
\end{align}
where $O_A$  and $O_B$ are Hermitian operators and $\ket{\psi}$ is an arbitrary state. This inequality can be proven by using the Cauchy–Schwarz inequality:
\begin{align}
    |\braket{\psi_1|\psi_2}| \leq \sqrt{\braket{\psi_1|\psi_1} \braket{\psi_2|\psi_2}}\,.
\end{align}
Replacing $\ket{\psi_1}$ by $\ket{O_A\psi}$ and $\ket{\psi_2}$ by $\ket{O_B\psi}$ in the Cauchy–Schwarz inequality, we can obtain Eq.~\eqref{eq:OABine} immediately. Let $O_A = \sum_{\mathbf{n},i} \frac{J^i(\mathbf{n})}{M^{\frac{d}{2}}\sqrt{2\hat{E}_{\mathbf{p}}}} \left[{\rm Re}(\hat{\epsilon}^s_i(\mathbf{p}))\cos(\mathbf{p}\cdot \mathbf{n})-{\rm Im}(\hat{\epsilon}^s_i(\mathbf{p}))\sin(\mathbf{p}\cdot \mathbf{n})\right]$ and $O_B = a^s_{\mathbf{p}}+a^{s\dagger}_{\mathbf{p}}$, the upper bound of $|\braket{\hat{H}^{S}_I(\mathbf{q},s)}|$ can be written as
\begin{align}
    |\braket{\hat{H}^{S}_I(\mathbf{q},s)}| &\leq \sqrt{\left\langle \left\{ \sum_{\mathbf{n},i} \frac{J^i(\mathbf{n})}{M^{\frac{d}{2}}\sqrt{2\hat{E}_{\mathbf{q}}}} \left[{\rm Re}(\hat{\epsilon}^s_i(\mathbf{q}))\cos(\mathbf{q}\cdot \mathbf{n})-{\rm Im}(\hat{\epsilon}^s_i(\mathbf{q}))\sin(\mathbf{q}\cdot \mathbf{n})\right]\right\}^2\right\rangle \braket{(a^s_{\mathbf{q}}+a^{s\dagger}_{\mathbf{q}})^2}}\nonumber \\
    &\leq \frac{4 d g M^{\frac{d}{2}}}{\sqrt{2\hat{E}_{\mathbf{q}}}} \sqrt{\braket{(a^s_{\mathbf{q}}+a^{s\dagger}_{\mathbf{q}})^2}} \leq \frac{4 d g M^{\frac{d}{2}}}{\sqrt{\hat{E}_{\mathbf{q}}}} \sqrt{2\braket{\hat{\mathcal{N}}_{\mathbf{q},s}}+1}\,.
\end{align}
Similarly, we can compute the upper bound of $\braket{\hat{H}^{A}_I(\mathbf{q},s)}$, and the result is the same:
\begin{align}
    \braket{\hat{H}^{A}_I(\mathbf{q},s)} \leq \frac{4 d g M^{\frac{d}{2}}}{\sqrt{\hat{E}_{\mathbf{q}}}} \sqrt{2\braket{\hat{\mathcal{N}}_{\mathbf{q},s}}+1}\,.
\end{align}
Finally, we have
\begin{align}
    \hat{E}_{\mathbf{q}} \braket{\hat{\mathcal{N}}_{\mathbf{p},s}} - \frac{8 d g M^{\frac{d}{2}}}{\sqrt{\hat{E}_{\mathbf{q}}}} \sqrt{2\braket{\hat{\mathcal{N}}_{\mathbf{q},s}}+1} < E + \frac{[3+4d^2(d-1)] g^2 M^{2d+2}}{8 \pi^2} + 2M^d  \sqrt{3+m^2+12mw+36w^2}\,.
\end{align}
From the above inequality, we can solve the upper bound of $\braket{\hat{\mathcal{N}}_{\mathbf{q},s}}$:
\begin{align}
    \braket{\hat{\mathcal{N}}_{\mathbf{q},s}} &\leq \frac{1}{2} \left\{\left[\frac{8dgM^{d/2}}{\hat{E}_{\mathbf{q}}^{3/2}}+\left(\frac{64 d^2 g^2 M^d}{\hat{E}_{\mathbf{q}}^3}+ \frac{[3+4 d^2 (d-1)] g^2 M^{2d+2}}{4 \pi^2 \hat{E}_{\mathbf{q}}} \right. \right. \right. \nonumber \\
    &\left. \left. \left. \quad + \frac{4M^d}{\hat{E}_{\mathbf{q}}}\sqrt{3+m^2+12mw+36w^2}+\frac{2E}{\hat{E}_{\mathbf{q}}}\right)^{\frac{1}{2}}\right]^2 -1\right\}\nonumber \\
    &\leq \frac{1}{2} \left\{\left[\frac{8 d g}{(2\pi)^{\frac{3}{2}}}  M^{\frac{d+3}{2}}+\left(\frac{8 d^2 g^2}{\pi^3} M^{d+3} +\frac{[3 + 4 d^2 (d-1)] g^2}{8\pi^3} M^{2d+3}\right. \right. \right. \nonumber\\ 
    &\left.\left.\left. \quad + \frac{2}{\pi}\sqrt{3+m^2+12mw+36w^2} M^{d+1} + \frac{E}{\pi} M \right)^{\frac{1}{2}}\right]^2 -1\right\} \equiv \max(\braket{\hat{\mathcal{N}}_{\mathbf{q},s}})\,.
\end{align}
Using Markov's inequality in Eq.~\eqref{eq:MKine}, we find that:
\begin{align}
    P(\hat{\mathcal{N}}_{\mathbf{q},s}\geq \Lambda) \leq \frac{\braket{\hat{\mathcal{N}}_{\mathbf{q},s}}}{\Lambda} \leq \frac{\max(\braket{\hat{\mathcal{N}}_{\mathbf{q},s}})}{\Lambda}\,.
\end{align}
We have $(d-1) M^d$ different operators $\hat{\mathcal{N}}_{\mathbf{q},s}$, so the $\max(\braket{\hat{\mathcal{N}}_{\mathbf{q},s}})/\Lambda$ should satisfy $\max(\braket{\hat{\mathcal{N}}_{\mathbf{q},s}})/\Lambda = \varepsilon_s/[(d-1)M^3]$ to make the total error bounded by $\varepsilon_s$. This gives us the final upper bound of the cutoff $\Lambda$:
\begin{align}
    \Lambda &\leq \frac{(d-1) M^d}{2\varepsilon_s} \left\{\left[\frac{8 d g}{(2\pi)^{\frac{3}{2}}}  M^{\frac{d+3}{2}}+\left(\frac{8 d^2 g^2}{\pi^3} M^{d+3} +\frac{[3 + 4 d^2 (d-1)] g^2}{8\pi^3} M^{2d+3}\right. \right. \right. \nonumber\\ 
    &\left.\left.\left. \quad + \frac{2}{\pi}\sqrt{3+m^2+12mw+36w^2} M^{d+1} + \frac{E}{\pi} M \right)^{\frac{1}{2}}\right]^2 -1\right\} \equiv \Lambda_{\max}\,.
\end{align}
Therefore, encoding the photon with given momentum $\mathbf{p}$ and polarization $r$ needs no more than $\log_2\Lambda_{\max}$ qubits. The total cost of qubits for encoding the gauge field is given by:
\begin{align}
    N^{\rm q,G}_{\rm CG} = (d-1) M^d \log_2 \Lambda_{\rm max}\,.
\end{align}

\section{Gate cost estimation of time evolution of CG QED}
\begin{figure}
    \centering
    \includegraphics[width=0.98\linewidth]{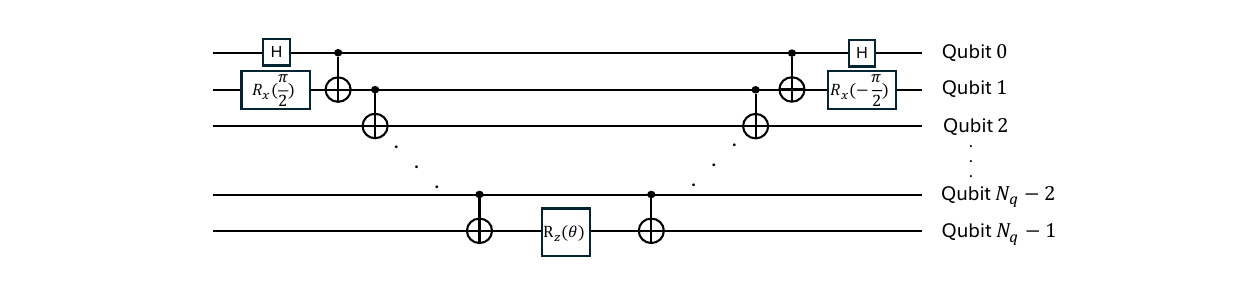}
    \caption{Quantum circuit of simulating $e^{-ih_\alpha \delta t/2}$, with $h_\alpha = C_\alpha \sigma^3_{{N_q}-1}...\sigma^3_{2}\sigma^2_{1}\sigma^1_{0}$. Here, we set the parameter $\theta = \delta t C_\alpha$. Analogous circuits can represent other evolution operators generated by $h_\alpha$.}
    \label{fig:cir}
\end{figure}

The Hamiltonian $H$ of an $N_q$ qubit system can be represented by the linear combination of Pauli strings $\sigma^{\mu_{N_q-1}}_{N_q-1} \otimes...\otimes \sigma^{\mu_0}_0$:
\bea\label{eq:deham}
    H = \sum_{\mu_0,...,\mu_{N_q-1}=0}^3 C_{\mu_{N_q-1},...,\mu_0} \sigma^{\mu_{N_q-1}}_{N_q-1} \otimes ... \otimes \sigma^{\mu_0}_0 \equiv \sum_{\alpha=1}^{\mathcal{M}_G} h_\alpha\,,
\eea
where $\mathcal{M}_G$ is the number of non-vanishing terms inside the summation $\sum_{\mu_0,...,\mu_{N_q-1}}$. According to the second-order Suzuki formula, the time evolution operator $e^{-iHt}$ can be approximated by:
\bea
    e^{-iHt} &\approx \left[\prod_{\alpha = \mathcal{M}_G}^{1} \exp\left(-i h_\alpha \frac{t}{2 N_T}\right) \prod_{\alpha=1}^{\mathcal{M}_G} \exp\left(-i h_\alpha \frac{t}{2 N_T}\right)+\mathcal{O}\left(\frac{t^3}{N_T^3}\right)\right]^{N_T} \nonumber\\
    &=\left[\prod_{\alpha = \mathcal{M}_G}^{1} \exp\left(-i h_\alpha \frac{\delta t}{2}\right) \prod_{\alpha=1}^{\mathcal{M}_G} \exp\left(-i h_\alpha \frac{\delta t}{2}\right)\right]^{N_T} +\mathcal{O}\left(\frac{t^3}{N_T^2}\right).
\eea
We have $\mathcal{O}(\varepsilon) = \mathcal{O}(t^3/N_T^2)$ for a given precision $\varepsilon$, so we can obtain $N_T = t^{3/2}/\varepsilon^{1/2}$. The gate cost of simulating $\exp(-i h_\alpha \delta t/2)$ is well known. For example, the quantum circuit in Fig.~\ref{fig:cir} can be used to simulate the evolution operator $e^{-ih_\alpha \delta t/2}$, with $h_\alpha = C_\alpha \sigma^3_{{N_q}-1}...\sigma^3_{2}\sigma^2_{1}\sigma^1_{0}$. This means that as long as we decompose the Hamiltonian into the form of Eq.~(\ref{eq:deham}), we can estimate the number of quantum gates required to simulate the time evolution operator. In the main text, we already know that the Hamiltonian of CG QED is given by $\hat{H} = \hat{H}_E+\hat{H}_B+\hat{H}_I+\hat{H}_V+\hat{H}_M+\hat{H}_W$. Next, we will give the specific expressions of the Hamiltonian after each term is decomposed into Eq.~(\ref{eq:deham}) form.

In the main text, the $\hat{H}_E+\hat{H}_B$ is given by
\bea
    \hat{H}_E + \hat{H}_B = \sum_{\mathbf{p}\not= 0}\sum_r \hat{E}_\mathbf{p} a^{r\dagger}_\mathbf{p} a^{r}_\mathbf{p} = \sum_{\mathbf{p}\not= 0}\sum_r \sum_{J=0}^{K-1} 2^J\left[\frac{1}{2} (I-\sigma^3_{J,\mathbf{p},r})\right]\,.
\eea
After dropping the identity operator, the time evolution operator $e^{-i(\hat{H}_E+\hat{H}_B)\delta t/2}$ can be decomposed as
\bea
    e^{-i(\hat{H}_E+\hat{H}_B)\delta t/2} = \prod_{\mathbf{p} \not= 0} \prod_r \prod_{J=0}^{K-1} \exp\left(i 2^{J-1} \sigma^3_{J,\mathbf{p},r} \delta t\right)\,,
\eea
which is nothing but the $R_z$ gate on a quantum computer. For $(d+1)$-dimensional space-time, $\prod_{\mathbf{p}}$ and $\prod_r$ have $M^d$ and $d-1$ different terms. Therefore, we need $N^{G,EB}_{\rm OQ} = (d-1)M^dK \approx (d-1)M^d \log_2\Lambda$ one-qubit gates to simulate $e^{-i(\hat{H}_E+\hat{H}_B)\delta t /2}$. We use $\approx$ here because $\Lambda = 2^K-1 \approx2^K$ when $2^K \gg 1$.

$e^{-i \hat{H}_I \delta t/2}$ is the most complex part in the time evolution operator. In the main text, the $\hat{H}_I$ is given by:
\bea\label{eq:HI_SM}
    \hat{H}_I = \sum_{\mathbf{n},i} \sum_{\mathbf{p}\not=0} \sum_{r} \frac{J^i(\mathbf{n})}{M^{\frac{3}{2}}\sqrt{2\hat{E}_{\mathbf{p}}}}\left[\hat{\epsilon}^r_i (\mathbf{p}) a^r_\mathbf{p} e^{i\mathbf{p}\cdot \mathbf{n}}+{\rm H.c.}\right]\,,
\eea
where the qubit form of $a^r_\mathbf{p}$ and $\sqrt{\hat{N}_{\mathbf{p},r}}$ are given by
\bea\label{eq:a_SM}
    a^{r}_{\mathbf{p}} = \left\{\sum_{J=0}^{K-1} \left[\sigma^+_{J,\mathbf{p},r} \left(\prod_{L=0}^{J-1}\sigma^-_{L,\mathbf{p},r}\right)\right]\right\} \sqrt{\hat{\mathcal{N}}_{\mathbf{p},r}}\,,
\eea
and
\bea\label{eq:sqrtN_SM}
    \sqrt{\hat{\mathcal{N}}_{\mathbf{p},r}} = \sum_{\mu_0,...,\mu_{K-1}=0,3} f_{\mu_{K-1},...,\mu_0} \sigma^{\mu_{K-1}}_{{K-1},\mathbf{p},r}\otimes ... \otimes \sigma^{\mu_0}_{0,\mathbf{p},r}\,.
\eea
The expression of coefficients $f_{\mu_{K-1},...,\mu_0}$ can be found in the main text. Putting Eq.~(\ref{eq:a_SM}) and Eq.~(\ref{eq:sqrtN_SM}) into Eq.~(\ref{eq:HI_SM}), we have
\bea
    \hat{H}_I =& \sum_{\mathbf{n},i} \sum_{\mathbf{p}\not=0} \sum_{r} \sum_{L=0}^{K-1} \sum_{\mu_{L+1},...,\mu_{K-1}=0,3}\frac{J^i(\mathbf{n})}{M^{\frac{3}{2}}\sqrt{2\hat{E}_{\mathbf{p}}}} \nonumber \\
    & \times \left\{\left[\Re(\hat{\epsilon}^r_i(\mathbf{p})) \cos(\mathbf{p}\cdot \mathbf{n})-\Im(\hat{\epsilon}^r_i(\mathbf{p})) \sin(\mathbf{p}\cdot \mathbf{n})\right] \mathcal{F}_{\mu_{K-1},...,\mu_{L+1}}\right. \nonumber \\
    &\left. \times \sigma^{\mu_{K-1}}_{K-1,\mathbf{p},r}...\sigma^{\mu_{L+1}}_{L+1,\mathbf{p},r}(\sigma^+_{L,\mathbf{p},r} \sigma^-_{L-1,,\mathbf{p},r}...\sigma^-_{0,\mathbf{p},r}+{\rm H.c.})\right.\nonumber\\
    &\left. +{\rm i}\left[\Re(\hat{\epsilon}^r_i(\mathbf{p})) \sin(\mathbf{p}\cdot \mathbf{n})+\Im(\hat{\epsilon}^r_i(\mathbf{p})) \cos(\mathbf{p}\cdot \mathbf{n})\right] \mathcal{F}_{\mu_{K-1},...,\mu_{L+1}}\right.\nonumber\\
    &\left. \times \sigma^{\mu_{K-1}}_{K-1,\mathbf{p},r}...\sigma^{\mu_{L+1}}_{L+1,\mathbf{p},r}(\sigma^+_{L,\mathbf{p},r} \sigma^-_{L-1,\mathbf{p},r}...\sigma^-_{0,\mathbf{p},r}-{\rm H.c.})\right\}\nonumber \\
\eea
where $\mathcal{F}_{\mu_{K-1},...,\mu_{L+1}}  = \sum_{\mu_0,...,\mu_L=0,3}(-1)^{\mu_L} f_{\mu_{K-1},...,\mu_0}$. From the above equation, we find that the $\hat{H}_I$ needs to divide into two parts, $\hat{H}_I = \hat{H}^S_I+\hat{H}^A_I\equiv \sum_{\mathbf{n},i}\sum_{\mathbf{p}\not=0}\sum_r[\hat{\mathcal{H}}^S_I(\mathbf{n},\mathbf{p},i,r)+\hat{\mathcal{H}}^A_I(\mathbf{n},\mathbf{p},i,r)]$:
\bea\label{eq:ham_HIS}
    \hat{H}^S_I &=\sum_{\mathbf{n},i} \sum_{\mathbf{p}\not=0} \sum_{r} \sum_{L=0}^{K-1} \sum_{\substack{\mu_0,...,\mu_L=1,2 \\ \mu_0+...+\mu_L-L-1={\rm even}}} \sum_{\mu_{L+1},...,\mu_{K-1}=0,3} \frac{2^{-L}J^i(\mathbf{n})}{M^{\frac{3}{2}}\sqrt{2\hat{E}_{\mathbf{p}}}} \nonumber \\
    &\quad \times \left[\Re(\hat{\epsilon}^r_i(\mathbf{p})) \cos(\mathbf{p}\cdot \mathbf{n})-\Im(\hat{\epsilon}^r_i(\mathbf{p})) \sin(\mathbf{p}\cdot \mathbf{n})\right] \mathcal{F}_{\mu_{K-1},...,\mu_{L+1}} \mathcal{G}^S_{\mu_{L},...,\mu_0}\sigma^{\mu_{K-1}}_{K-1,\mathbf{p},r}...\sigma^{\mu_{L+1}}_{L+1,\mathbf{p},r}\sigma^{\mu_L}_{L,\mathbf{p},r} ...\sigma^{\mu_0}_{0,\mathbf{p},r}\,,
\eea
where $\mathcal{G}^S_{\mu_L,...,\mu_0} = (-1)^{\mu_L-1}(-1)^{\left(-L-1+\sum_{J=0}^{L}\mu_J\right)/2}$, and the $\hat{H}^A_I$ can be written as:
\bea\label{eq:ham_HIA}
    \hat{H}^A_I &=\sum_{\mathbf{n},i} \sum_{\mathbf{p}\not=0} \sum_{r} \sum_{L=0}^{K-1} \sum_{\substack{\mu_0,...,\mu_L=1,2 \\ \mu_0+...+\mu_L-L-1={\rm odd}}} \sum_{\mu_{L+1},...,\mu_{K-1}=0,3} \frac{2^{-L}J^i(\mathbf{n})}{M^{\frac{3}{2}}\sqrt{2\hat{E}_{\mathbf{p}}}} \nonumber \\
    &\quad \times \left[\Re(\hat{\epsilon}^r_i(\mathbf{p})) \sin(\mathbf{p}\cdot \mathbf{n})+\Im(\hat{\epsilon}^r_i(\mathbf{p})) \cos(\mathbf{p}\cdot \mathbf{n})\right] \mathcal{F}_{\mu_{K-1},...,\mu_{L+1}} \mathcal{G}^A_{\mu_{L},...,\mu_0}\sigma^{\mu_{K-1}}_{K-1,\mathbf{p},r}...\sigma^{\mu_{L+1}}_{L+1,\mathbf{p},r}\sigma^{\mu_L}_{L,\mathbf{p},r} ...\sigma^{\mu_0}_{0,\mathbf{p},r}\,,
\eea
where $\mathcal{G}^A_{\mu_L,...,\mu_0} = (-1)^{\mu_L-1}(-1)^{\left(-L-2+\sum_{J=0}^{L}\mu_J\right)/2}$. The electric current operator $J^i(\mathbf{n}) = g\bar{\psi}(\mathbf{n})\gamma^i \psi(\mathbf{n})$ in Eq.~(\ref{eq:ham_HIS}) and Eq.~(\ref{eq:ham_HIA}) can be mapped to qubits by Jordan-Wigner transformation:
\bea\label{eq:JWJi}
    &\bar{\psi}(\mathbf{n}) \gamma^1 \psi(\mathbf{n}) = \frac{g}{2}\sum_{h=0,2} (-1)^{\frac{h}{2}} \left(\sigma^1_{\mathbf{n},h+1}\sigma^1_{\mathbf{n},h+2}+\sigma^2_{\mathbf{n},h+1}\sigma^2_{\mathbf{n},h+2}\right)\,\nonumber \\
    &\bar{\psi}(\mathbf{n}) \gamma^2 \psi(\mathbf{n}) = -\frac{g}{2}\sum_{h=0,2} (-1)^{\frac{h}{2}} \left(\sigma^1_{\mathbf{n},h+1}\sigma^2_{\mathbf{n},h+2}-\sigma^2_{\mathbf{n},h+1}\sigma^1_{\mathbf{n},h+2}\right)\,\nonumber\\
    &\bar{\psi}(\mathbf{n}) \gamma^3 \psi(\mathbf{n}) = \frac{g}{2} \sum_{h=0,2} (-1)^{\frac{h}{2}} \left(\sigma^3_{\mathbf{n},h+1}-\sigma^3_{\mathbf{n},h+2}\right)\,,
\eea
where $h=0$ and $h=2$ come from the contribution of left and right-handed fermion fields, respectively. In the case of $d=2,3$, the Dirac gamma matrices are given by:
\bea
    \gamma^0=\left[\begin{array}{cc}
    0 & I \\
    I & 0
    \end{array}\right]\,,  \ \gamma^i=\left[\begin{array}{cc}
    0 & \sigma^i \\
    -\sigma^i & 0
    \end{array}\right]\,.
\eea
Now, we are ready to estimate the gate cost to simulate the evolution operator $e^{-i(\hat{H}^S_I+\hat{H}^A_I)\delta t/2} \approx e^{-i\hat{H}^S_I \delta t/2}e^{-i\hat{H}^A_I \delta t/2}$. To achieve this, we need to calculate the length of the qubit string in $\hat{H}^S_I$ and $\hat{H}^A_I$. The evolution operator generated by length $L$ Pauli string needs to be represented by $2(L-1)$ CNOT gates in the ideal case. The lengths of Pauli string $J^i(\mathbf{n})\sigma^{\mu_L}_{L}...\sigma^{\mu_0}_0$ are $L+3$ for $i=1,2$ and $L+2$ for $i=3$. However, the length of $\sigma^{\mu_{K-1}}_{K-1}...\sigma^{\mu_{L+1}}_{L+1}$ needs to be treated more carefully because the identity operator will not increase the length of the Pauli string. There are $C^I_{K-L-1}$ terms that have total Pauli string length $L+3+I$ in the summation $\sum_{\mu_{L+1},...,\mu_{K-1}}$, while $\sum_{\substack{\mu_0,...,\mu_L=1,2 \\ \mu_0+...+\mu_L-L-1={\rm even (odd)}}}$ contribute $2^L$ terms. Therefore, the CNOT gate cost of simulating the evolution $\exp\left[-i \left(\hat{\mathcal{H}}^S_I(\mathbf{n},\mathbf{p},i,r)+\hat{\mathcal{H}}^A_I(\mathbf{n},\mathbf{p},i,r)\right)\delta t/2\right]$ is:
\bea\label{eq:CNOTHI1}
	&\left[\sum_{L=0}^{K-1} \sum_{I=0}^{K-L-1} 2(L+2+I) C^I_{K-L-1} 2^L\right]\times 2^3 = K(3K+5)2^{K+1} \approx 2\Lambda (\log_2\Lambda) (3\log_2\Lambda+5)\ {\rm if}\ i=1,2\,,\nonumber\\
    &\left[\sum_{L=0}^{K-1} \sum_{I=0}^{K-L-1} 2(L+1+I) C^I_{K-L-1} 2^L\right]\times 2^3 \approx 2\Lambda (\log_2\Lambda) (3\log_2\Lambda+1)\ {\rm if}\ i=3\,.
\eea
The Eq.~(\ref{eq:CNOTHI1}) dose not depend on $\mathbf{n}$, $\mathbf{p}$, and $r$, so the total CNOT gate cost of simulating $e^{-i\hat{H}_I \delta t/2}$ is
\bea
	N^{G,I}_{\rm CNOT} = 2(d-1)M^{2d} \left[d\Lambda (\log_2\Lambda) (3\log_2\Lambda+5)-4\Lambda\log_2\Lambda\right]\,.
\eea
Similarly, we can estimate the single qubit gate cost of simulating $e^{-i\hat{H}_I \delta t/2}$:
\bea
	N^{G,I}_{\rm OQ} = 4(d-1)M^{2d} \left[d \Lambda (\log_2 \Lambda) (\log_2\Lambda + 5) - 4\Lambda \log_2\Lambda\right]\,.
\eea

Then we need to consider the complexity of simulating $e^{-iH_V \delta t/2}$. In the main text, $H_V$ is given by
\bea
	\hat{H}_V = \frac{1}{2}\sum_{\mathbf{n},\mathbf{m}} \sum_{\mathbf{p}\not=0} \frac{J^0(\mathbf{n})J^0(\mathbf{m})}{\hat{E}_{\mathbf{p}}^2}e^{-i\mathbf{p}\cdot(\mathbf{n}-\mathbf{m})}
\eea
After the Jordan-Wigner transformation, we can translate $\hat{H}_V$ to the linear combination of Pauli strings:
\bea
	\hat{H}_V = -\frac{g^2 N_s}{8}\sum_{\mathbf{n},\alpha} \sum_{\mathbf{m}} \sum_{\mathbf{p}\not=0} \frac{1}{\hat{E}_{\mathbf{p}}} e^{-i\mathbf{p}\cdot (\mathbf{n}-\mathbf{m})} \sigma^3_{\mathbf{n},\alpha}+\frac{g^2}{8} \sum_{\substack{\mathbf{n}\not= \mathbf{m} \\ \alpha\not=\beta}} \sum_{\mathbf{p}\not=0} \frac{1}{\hat{E}_{\mathbf{p}}} e^{-i\mathbf{p}\cdot (\mathbf{n}-\mathbf{m})} \sigma^3_{\mathbf{n},\alpha} \sigma^3_{\mathbf{m},\beta}\,.
\eea
In the above equation, we dropped the constant term in the Hamiltonian $\hat{H}_V$. There are $N_s M^{2d}$ and $N_s(N_s-1)M^d(M^d-1)$ of one-Pauli and ``$zz$" two-Pauli terms inside the Hamiltonian $\hat{H}_V$. Therefore, the one and two-qubit gate cost of simulating $e^{-iH_V \delta t/2}$ are
\bea
	&N^{G,V}_{\rm OQ} = N_s M^{2d}\,, \nonumber\\
	&N^{G,V}_{\rm CNOT} = 2N_s M^d (N_s M^d-1)\,.
\eea

Kinetic and mass terms of the fermion field are given by:
\bea
    \hat{H}_M = \sum_{\mathbf{n}} \bar{\psi}(\mathbf{n})\left[-{\rm i}\sum_i \gamma^i \frac{\psi(\mathbf{n}+\hat{i})-\psi(\mathbf{n}-\hat{i})}{2}+m\bar{\psi}(\mathbf{n})\psi(\mathbf{n})\right]\,.
\eea
After the Jordan-Wigner transformation, the $\hat{H}_M$ is given by
\bea\label{eq:Hm_pau}
    \hat{H}_M =& \sum_{\mathbf{n}} \sum_{h=0,2}\sum_{i=1}^{d} \sum_{s_1,s_2=1}^2 -\frac{(-1)^{\frac{h}{2}}\sigma^i_{s_1 s_2}}{4} \left(\prod_{\gamma=s_1+1+h}^{N_s} \sigma^3_{\mathbf{n},\gamma}\right)\left[\prod_{l=1}^{M^{i-1}-1}\left(\prod_{\gamma=1}^{N_s} \sigma^3_{\mathbf{m}(l),\gamma}\right)\right]\left(\prod_{\gamma=1}^{s_2-1+h}\sigma^3_{\mathbf{n}+\hat{i},\gamma}\right)\nonumber \\
    &\times \left(\sigma^1_{\mathbf{n},s_1+h}\sigma^2_{\mathbf{n}+\hat{i},s_2+h}-\sigma^2_{\mathbf{n},s_1+h}\sigma^1_{\mathbf{n}+\hat{i},s_2+h}\right) - \frac{m}{2} \sum_{\mathbf{n}} \sum_{s=1}^2 \left(\sigma^1_{\mathbf{n},s} \sigma^3_{\mathbf{n},s+1} \sigma^1_{\mathbf{n},s+2}+\sigma^2_{\mathbf{n},s} \sigma^3_{\mathbf{n},s+1} \sigma^2_{\mathbf{n},s+2}\right)\,,
\eea
In the Eq.~(\ref{eq:Hm_pau}), $\mathbf{m}(l)$ is defined as
\bea
    &m_x(l) = (n_x+l)\mod M \nonumber\\
    &m_y(l) = \left[n_y+\frac{n_x+l-m_x(l)}{M}\right] \mod M \nonumber\\
    &m_z(l) = n_z+\left[n_y+\frac{n_x+l-m_x(l)}{M}-m_y(l)\right] \mod M\,.
\eea
Then, we can obtain the one-qubit and CNOT gate cost of simulating $e^{-i\hat{H}_M \delta t/2}$ in $d=2$ and $d=3$:
\bea
    &N^{G,M}_{\rm OQ}(d=2) = 72 M^2\,,\nonumber\\
    &N^{G,M}_{\rm OQ}(d=3) = 104 M^3\,,\nonumber\\
    &N^{G,M}_{\rm CNOT}(d=2) = 16 N_s M^2(1+M) + 8M^2\,\nonumber\\
    &N^{G,M}_{\rm CNOT}(d=3) = 16 N_s M^3(1+M+M^2) + 8M^3\,.
\eea

At last, we discuss the complexity of simulating $e^{-i\hat{H}_W \delta t/2}$. The Hamiltonian $\hat{H}_W$ is given by
\bea
    \hat{H}_W = \sum_{\mathbf{n}} -\frac{w}{2} \bar{\psi}(\mathbf{n}) \hat{\Delta} \psi(\mathbf{n}) = \sum_{\mathbf{n}} \sum_{i=1}^d -\frac{w}{2a^2} \bar{\psi}(\mathbf{n}) \left[ \psi(\mathbf{n}+\hat{i})+\psi(\mathbf{n}-\hat{i})-2\psi(\mathbf{n})\right]\,.
\eea
After the Jordan-Wigner transformation, the Wilson term $\hat{H}_W$ can be written as
\bea
    \hat{H}_W =& \sum_{\mathbf{n}} \sum_{i=1}^d \sum_{s=1}^2 \frac{w}{4a^2} \left(\prod_{\gamma=s+1}^{N_s} \sigma^3_{\mathbf{n},\gamma}\right)\left[\prod_{l=1}^{M^{i-1}-1}\left(\prod_{\gamma=1}^{N_s} \sigma^3_{\mathbf{m}(l),\gamma}\right)\right]\left(\prod_{\gamma=1}^{s+1} \sigma^3_{\mathbf{n},\gamma}\right)\left(\sigma^1_{\mathbf{n},s}\sigma^1_{\mathbf{n}+\hat{i},s+2} + \sigma^2_{\mathbf{n},s}\sigma^2_{\mathbf{n}+\hat{i},s+2}\right)\nonumber \\
    &+ \sum_{\mathbf{n}} \sum_{i=1}^d \sum_{s=1}^2 \frac{w}{4a^2} \left(\prod_{\gamma=s+3}^{N_s} \sigma^3_{\mathbf{n},\gamma}\right)\left[\prod_{l=1}^{M^{i-1}-1}\left(\prod_{\gamma=1}^{N_s} \sigma^3_{\mathbf{m}(l),\gamma}\right)\right]\left(\prod_{\gamma=1}^{s-1} \sigma^3_{\mathbf{n},\gamma}\right)\left(\sigma^1_{\mathbf{n},s+2}\sigma^1_{\mathbf{n}+\hat{i},s} + \sigma^2_{\mathbf{n},s+2}\sigma^2_{\mathbf{n}+\hat{i},s}\right)\nonumber \\
    &- \frac{w}{2a^2} \sum_{\mathbf{n}} \sum_{s=1}^2 \left(\sigma^1_{\mathbf{n},s} \sigma^3_{\mathbf{n},s+1} \sigma^1_{\mathbf{n},s+2}+\sigma^2_{\mathbf{n},s} \sigma^3_{\mathbf{n},s+1} \sigma^2_{\mathbf{n},s+2}\right)\,.
\eea
Then, the one-qubit and CNOT gate cost of simulating $e^{-i\hat{H}_W \delta t/2}$ is in $d=2$ and $d=3$ is given by
\bea
    &N^{G,W}_{\rm OQ}(d=2) = 72 M^2\,,\nonumber\\
    &N^{G,W}_{\rm OQ}(d=3) = 104 M^3\,,\nonumber\\
    &N^{G,W}_{\rm CNOT}(d=2) = 16 N_s M^2(1+M) + 8M^2\,\nonumber\\
    &N^{G,W}_{\rm CNOT}(d=3) = 16 N_s M^3(1+M+M^2) + 8M^3\,,
\eea
which are the same with the gate cost of simulating $e^{-i\hat{H}_M \delta t/2}$.

Finally, the total single and CNOT gate cost of simulating $e^{-i\hat{H}t}$, which are denoted as $N_{G,1}$ and $N_{G,2}$, depend on the Trotter step $N_T$. As discuss before, for given $t$ and precision $\varepsilon$, $N_T = t^{3/2}/\varepsilon^{1/2}$, so we have
\bea
    &N^{G}_{\rm CG,OQ} = \frac{2 t^{3/2}}{\varepsilon^{1/2}}\left(N^{G,EB}_{\rm OQ}+N^{G,I}_{\rm OQ}+N^{G,V}_{\rm OQ}+N^{G,M}_{\rm OQ}+N^{G,W}_{\rm OQ}\right)\,,\nonumber\\
    &N^{G}_{\rm CG,CNOT} = \frac{2t^{3/2}}{\varepsilon^{1/2}}\left(N^{G,EB}_{\rm CNOT}+N^{G,I}_{\rm CNOT}+N^{G,V}_{\rm CNOT}+N^{G,M}_{\rm CNOT}+N^{G,W}_{\rm CNOT}\right)\,.
\eea
The cost of quantum gates depends on $M$, $\Lambda$, $\varepsilon$, and $t$ polynomially.

\end{document}